\documentstyle[preprint,tighten,aps,floats,psfig,epsfig]{revtex}

\def\o{\omega}
\def\e{\epsilon}

\begin{document}
\draft
\title{
Spin models with random  anisotropy and reflection symmetry
}
\author{Pasquale Calabrese,$^{1,2}$ Andrea Pelissetto,$^3$ 
Ettore Vicari$\,^4$ }
\address{$^1$ Scuola Normale Superiore and  INFN, Piazza dei Cavalieri 7,
 I-56126 Pisa, Italy.}
\address{$^2$ 
Theoretical Physics, University of Oxford,
1 Keble Road, Oxford OX1 3NP, United Kingdom.}
\address{$^3$ Dipartimento di Fisica dell'Universit\`a di Roma I
and INFN, I-00185 Roma, Italy.}
\address{$^4$
Dipartimento di Fisica dell'Universit\`a 
and INFN, 
Via Buonarroti 2, I-56127 Pisa, Italy.
{\bf e-mail: \rm 
{\tt calabres@df.unipi.it},
{\tt Andrea.Pelissetto@roma1.infn.it},
{\tt vicari@df.unipi.it}.
}}

\date{\today}

\maketitle

\begin{abstract}
We study the critical behavior of a
general class of cubic-symmetric spin systems in which disorder preserves the 
reflection symmetry $s_a\to -s_a$, $s_b\to s_b$ for $b\not= a$.
This includes spin models in the presence
of random cubic-symmetric anisotropy with probability distribution 
vanishing outside the lattice axes. 
Using nonperturbative arguments we show the existence
of a stable fixed point corresponding to the random-exchange 
Ising universality class. The field-theoretical renormalization-group flow is
investigated in the framework of a fixed-dimension
expansion in powers of appropriate quartic couplings,
computing the corresponding $\beta$-functions to five loops.
This analysis shows that the random Ising fixed point is the only stable 
fixed point that is accessible from the relevant parameter region. 
Therefore, if the system undergoes a continuous transition, 
it belongs to the random-exchange Ising universality class. 
The approach to the asymptotic critical behavior is controlled by
scaling corrections with exponent $\Delta = - \alpha_r$, 
where $\alpha_r\simeq -0.05$ is the specific-heat exponent of the
random-exchange Ising model. 
\end{abstract}



\section{Introduction and summary}
\label{intro}

The critical behavior of magnetic systems in the presence of quenched disorder 
has been the subject of extensive theoretical and experimental study.
An important class of systems is formed by amorphous alloys 
of rare earths with aspherical electron distribution and transition metals,
for instance TbFe$_2$ and YFe$_2$. These systems are modeled 
\cite{HPZ-73,revram} by the Heisenberg model with random uniaxial 
single-site anisotropy, or, in short, by the random-anisotropy model (RAM)
\begin{equation}
{\cal H}_{\rm RAM} =- J \sum_{\langle xy \rangle} \vec{s}_x \cdot \vec{s}_y-
D_0\sum_x(\vec{u}_x \cdot \vec{s}_x)^2,
\label{HRAM}
\end{equation}
where $\vec{s}_x$ is an $M$-component spin variable, $\vec{u}_x$ is a unit 
vector describing the local (spatially uncorrelated) random anisotropy, 
and $D_0$ the anisotropy strength. In amorphous alloys the distribution of 
$\vec{u}_x$ is usually taken to be isotropic, since in the absence of 
crystalline order there is no preferred direction. On the other hand, 
in polycrystalline materials, for instance in the 
Laves-phase intermetallic  (Dy$_x$Y$_{1-x}$)Al$_2$ compounds studied in 
Refs.~\cite{GSMA-90,MAGSRJC-93}, the distribution of 
$\vec{u}_x$ is expected to have only the lattice symmetry.

The critical behavior, and in particular the nature of the low-temperature
phase, of a generic system with random anisotropy 
depends on the probability distribution of the random vector $\vec{u}_x$.
In the isotropic case, i.e. when  the probability
distribution is uniformly weighted over the $(M-1)$-dimensional sphere,
the Imry-Ma argument \cite{IM-73,PPR-78} forbids the appearance of a 
low-temperature phase with nonvanishing magnetization for $d<4$.
This still allows the presence of a finite-temperature transition
with a low-temperature phase in which correlation functions decay
algebraically, as it happens in the two-dimensional XY model. 
Such a behavior has been predicted for the RAM with isotropic distribution 
in Ref.~\cite{AP-80} and it has been recently supported by a $4-\epsilon$ study 
\cite{Feldman-99,Feldman-01} using the functional renormalization group (RG) 
\cite{Fisher-85}. On the other hand, standard field-theoretical 
perturbative approaches do 
not find any evidence for a critical behavior with long-range correlations
\cite{Aharony-75,MG-82,DFH-01a,CPV-unpub}. While experiments have not yet
found evidence of low-temperature quasi long-range order, 
numerical simulations seem to confirm the picture of 
Refs.~\cite{AP-80,Feldman-99,Feldman-01}, but are still contradictory
as far as universality and
behavior in the strong-anisotropy regime are concerned
\cite{Fisch-98,Itakura-03}.
For these reasons, the critical behavior of the RAM can still
be considered as an open problem.  

The above arguments do not apply to 
spin models with the discrete anisotropic distribution
introduced in Ref.~\cite{Aharony-75}, in which the vector 
$\vec{u}_x$ points only along one of the $M$ 
lattice axes, i.e. it has the probability distribution 
\begin{equation}
P_c(\vec{u})=\frac{1}{2M}\sum_{a=1}^M[\delta^{(M)}(\vec{u}-\hat{x}_a)+
\delta^{(M)}(\vec{u}+\hat{x}_a)], 
\label{Cdistr}
\end{equation}
where $\hat{x}_a$ is a unit vector which points in the positive $a$ direction.
This model, which we shall call random cubic anisotropic model (RCAM), 
should have a standard order-disorder transition: 
The random discrete cubic anisotropy should stabilize a low-temperature 
phase with long-range ferromagnetic order.
On the basis of two-loop calculations in field-theoretical frameworks,
it has been argued \cite{DFH-01,HBDFFY-03} that the transition
belongs to the universality class of the random-exchange Ising model (REIM) 
for any number $M$ of components.

In this paper we study the critical properties
of the three-dimensional RCAM.
We consider the field-theoretical approach 
based on the Landau-Ginzburg-Wilson
$\varphi^4$ Hamiltonian \cite{Aharony-75} 
\begin{eqnarray}
{\cal H}_{\rm LGW} = \int d^dx \left
\{\sum_{i,a} \frac{1}{2} \left[ (\partial_\mu \phi_{ai})^2 + r \phi_{ai}^2 
\right]
+\frac{1}{4!} \sum_{ijab} (u_0 + v_0\delta_{ij} +
w_0\delta_{ab}+y_0\delta_{ij}\delta_{ab} ) \phi_{ai}^2\phi_{bj}^2
\right\},
\label{LGWCRAM} 
\end{eqnarray}
where $a,b=1,\ldots M$ and $i,j=1,\ldots N$. 
In the limit $N\rightarrow 0$ 
the Hamiltonian (\ref{LGWCRAM}) 
is expected to describe the critical behavior of the RCAM for $M$-component 
spins. Using nonperturbative arguments, we show
that the field theory with Hamiltonian (\ref{LGWCRAM}) has two stable fixed 
points (FP's). One of them belongs to the REIM universality class
while the other corresponds to the O($N$) model in the limit $N\rightarrow 0$,
the so-called self-avoiding-walk universality class.
Then, we investigate the RG flow for the model with 
Hamiltonian (\ref{LGWCRAM}) in the framework of a fixed-dimension
expansion in powers of appropriate  zero-momentum quartic couplings.
We compute the corresponding Callan-Symanzik $\beta$-functions to five loops.
Their analysis shows that the only accessible stable FP from the 
region of parameters relevant for the RCAM  is the REIM FP. 
This implies that
the critical behavior of the RCAM (when the parameters allow a continuous
transition) belongs to the REIM universality class,
whose critical exponents are \cite{CMPV-03}
$\nu_r=0.683(3)$, $\alpha_r=-0.049(9)$, $\eta_r=0.035(2)$, etc.
The approach to the REIM scaling behavior 
is characterized by very slowly decaying scaling corrections
proportional to $t^\Delta$ with $\Delta=-\alpha_r\approx 0.05$,
which is much smaller than the scaling-correction
exponent of the REIM, which is \cite{BFMMPR-98,CPPV-03} $\Delta_r\approx 0.25$.
Our results fully confirm and put on a firmer
ground the conclusions of Refs.~\cite{DFH-01,HBDFFY-03}
based on two-loop perturbative calculations.

It is important to note that our results are specific of 
distributions that vanish everywhere outside the lattice axes, 
such as the one given in Eq.~(\ref{Cdistr}).
Indeed, generic cubic-symmetric distributions $P(\vec{u})$, and in particular
the isotropic one, give rise to an additional quartic 
term that should be added to the effective Hamiltonian 
(\ref{LGWCRAM}), i.e.
\begin{equation}
z_0 \; \sum_{ijab} \phi_{ai} \phi_{bi} \phi_{aj} \phi_{bj}\; .
\end{equation}
The REIM FP is unstable with respect to this perturbation.
We shall evaluate the corresponding crossover exponent, finding
$\phi_z=0.79(4)$. 
Therefore, even small differences from the discrete distribution
$P_c(\vec{u})$ cause a crossover to  a different critical behavior.
Nonetheless, when $P_c(\vec{u})$ turns out to be a good effective 
approximation---this might be the case in some crystalline 
cubic-symmetric random-anisotropy systems---REIM
critical behavior may be observed in a preasymptotic region.

The general Landau-Ginzburg-Wilson Hamiltonian (\ref{LGWCRAM}) can also be 
recovered by considering systems with cubic anisotropy such that disorder
preserves the symmetry $s_{x,a}\to - s_{x,a}$, $s_{x,b} \to s_{x,b}$, 
$b \not= a$. A general Hamiltonian with this property is given by
\begin{equation}
{\cal H} =- J \sum_{\langle xy \rangle} \vec{s}_x \cdot \vec{s}_y
- K \sum_x \sum_a {s}_{x,a}^4
- D_0\sum_x \sum_a {q}_{x,a} {s}_{x,a}^2,
\label{HRCAM-gen}
\end{equation}
where $s_x^{\, 2} = 1$ and 
$\vec{q}_x$ is a random vector with a probability distribution 
that is invariant under the interchange $q_a\leftrightarrow q_b$.
The exact reflection symmetry at fixed disorder---this symmetry
is not present in generic models of type (\ref{HRAM})---is the 
key property that allows the 
RCAM and the more general class of models (\ref{HRCAM-gen}) to have 
a standard order-disorder transition with a low-temperature magnetized phase.

The paper is organized as follows.
In Sec.~\ref{sec1} we apply the replica method to the $\varphi^4$ theory
corresponding to models (\ref{HRAM}) and (\ref{HRCAM-gen}), 
determining the corresponding $\phi^4$ 
Hamiltonians that are the basis of the field-theoretical approach.
In Sec.~\ref{sec2} we discuss some general properties of 
the theory (\ref{LGWCRAM}). In particular, we discuss the crossover behavior 
when randomness is weak, and we prove that the 
REIM FP is stable by evaluating its stability eigenvalues.
In Sec.~\ref{sec3} we investigate the RG flow by computing
and analyzing the five-loop fixed-dimension expansion
of the $\beta$-functions associated with the zero-momentum quartic couplings.
In App.~\ref{quadrop} we report a six-loop  calculation 
of the RG dimensions of the bilinear operators
in cubic-symmetric models that are used
in the discussion of the stability of the FP's.
App.~\ref{betaided} reports the proof of some identities used in the paper.

\section{Effective $\mathbf{\Phi}^4$ Hamiltonians}
\label{sec1}

The mapping of the RAM Hamiltonian (\ref{HRAM}) 
to an effective translationally-invariant $\phi^4$ Hamiltonian 
was originally discussed in Ref.~\cite{Aharony-75}.
In order to replace fixed-length spins with uncostrained variables,
one performs a Hubbard-Stratonovich transformation. 
Then, for the purpose of studying the critical behavior
one considers the continuum limit of the resulting Hamiltonian and 
truncates its potential to fourth order.
This leads to an effective continuum $\varphi^4$ Hamiltonian for
an $M$-component real field $\varphi_a$ 
\begin{eqnarray}
{\cal H}_{\varphi^4} = \int d^dx \left[
\frac{1}{2} (\partial_\mu \vec{\varphi})^2 + \frac{1}{2} r \vec{\varphi}^{\,2} 
- D ( \vec{u}\cdot \vec{\varphi})^2 +\frac{1}{4!} v_0 (\vec{\varphi}^{\,2})^2
\right],
\label{heff} 
\end{eqnarray}
where $\vec{u}_x$ is an external spatially uncorrelated
vector field with parity-symmetric distribution $P(\vec{u})$ and $D$ is 
proportional to $D_0$.
We relax here the condition $\vec{u}^{\, 2}_x = 1$ and require only that 
$\langle \vec{u}^{\, 2}_x \rangle = 1$, thereby fixing the normalization of $D$.
Using the standard replica trick it is possible to
replace the quenched average with an annealed one.  The system is replaced by 
$N$ noninteracting copies with annealed disorder.  Then, by integrating over 
disorder, one obtains the following effective Hamiltonian
\begin{equation}
H_{\rm repl} = 
\int d^dx \left[
\sum_{i,a}  \frac{1}{2} (\partial_\mu \phi_{ai})^2 + 
\frac{1}{2} r \sum_{ia}   \phi_{ai}^2 
+\frac{1}{4!} v_0 \sum_{ijab} \delta_{ij} \phi_{ai}^2\phi_{bj}^2 +
R(\phi)\right]\, , 
\label{hrepl} 
\end{equation}
where $a,b=1,\ldots M$, $i,j=1,\ldots N$, and
\begin{equation}
R(\phi) = - {\rm ln} \int d^N u \, P(\vec{u})\, {\rm exp} 
\left( D \sum_{iab} u_a u_b  \phi_{ai}\phi_{bi}\right).
\end{equation}
In the limit $N\rightarrow 0$ the Hamiltonian (\ref{hrepl}) is equivalent
to the Hamiltonian (\ref{heff}) with quenched disorder.
The expansion in powers of the field $\phi$ can be expressed
in terms of the moments of the distribution $P(\vec{u})$,
\begin{equation}
M_{a_1 a_2 ...a_k} \equiv
\int d^N u \, P(\vec{u})\,  u_{a_1} u_{a_2} ...u_{a_k}\; .
\end{equation} 
One obtains
\begin{eqnarray}
&&H_{\rm repl} = 
\int d^dx \left[
 \frac{1}{2} \sum_{ia} (\partial_\mu \phi_{a,i})^2 + 
\frac{1}{2} r \sum_{ia} \phi_{a,i}^2 
+\frac{1}{4!} v_0 \sum_{iab} \phi_{ai}^2\phi_{bi}^2   \right.
\nonumber
\\
&& \left.
- D \sum_{iab} M_{ab} \phi_{ai} \phi_{bi} 
+ \frac{1}{2} D^2 ( \sum_{iab} M_{ab} \phi_{ai} \phi_{bi} )^2
- \frac{1}{2}  D^2 \sum_{ijabcd} M_{abcd} \phi_{ai} \phi_{bi} \phi_{cj} \phi_{dj}
+ O(\phi^6) \right] \; .
\label{hrepl2} 
\end{eqnarray}
Let us consider the case in which all field components become critical at $T_c$.
This is achieved if the distribution $P(\vec{u})$ is such that
\begin{equation}
M_{ab} = {1\over M} \delta_{ab}.
\end{equation}
This condition is satisfied if $P(u)$ is cubic symmetric.
Under this further assumption, the fourth moment $M_{abcd}$ can be written as
\begin{equation}
M_{abcd} = A \left( \delta_{ab}\delta_{cd} + 
\delta_{ac}\delta_{bd} + \delta_{ad}\delta_{bc} \right) + 
B \delta_{abcd}\; ,
\label{mabcd}
\end{equation}
where $A$ and $B$ are parameters depending on the distribution that satisfy
the Cauchy inequalities $A(M+2)+B\ge 1/M$ and $3 A + B\ge 1/M^2$. It
follows that 
\begin{eqnarray}
&&H_{\rm repl} = 
\int d^dx \left[ \vphantom{{D^2\over 2M^2}}
\frac{1}{2} \sum_{ia} (\partial_\mu \phi_{ai})^2 + 
\frac{1}{2} (r-D/M)\sum_{ia} \phi_{ai}^2 
+\frac{1}{4!} v_0 \sum_{iab}\phi_{ai}^2\phi_{bi}^2  + \right.
\nonumber \\
&& \left.
+ {D^2\over 2M^2} (1-M^2 A) (\sum_{ia} \phi_{ai}^2)^2
- A D^2 \sum_{ijab} \phi_{ai} \phi_{bi} \phi_{aj} \phi_{bj}
- {B D^2\over2} \sum_{ija} \phi_{ai}^2 \phi_{aj}^2
+ O(\phi^6) \right] .
\label{hrepl3} 
\end{eqnarray}
In conclusion, for generic cubic-symmetric distributions $P(\vec{u})$ 
the Hamiltonian that should describe the critical behavior of 
the corresponding RAM is 
\begin{eqnarray}
{\cal H} = && \int d^dx \left\{
\frac{1}{2} \sum_{ia} (\partial_\mu \phi_{ai})^2 + 
\frac{1}{2} r \sum_{ia}  \phi_{ai}^2 +  \right.
\nonumber \\
&& \left.
+\frac{1}{4!} \sum_{ijab} \Bigl[ (u_0 + v_0\delta_{ij} +
w_0\delta_{ab}+y_0\delta_{ij}\delta_{ab} ) \phi_{ai}^2\phi_{bj}^2
+ z_0 \phi_{ai} \phi_{bi} \phi_{aj} \phi_{bj} \right\}, 
\label{LGWgen} 
\end{eqnarray}
where the term proportional to $y_0$ has been added because it is 
generated by RG iterations whenever $w_0 \not = 0$. It should be noticed that 
such a term arises naturally if one considers that, if the system is only
cubic symmetric, quartic single-ion terms must be included. In this case
it is natural to consider 
\begin{equation}
{\cal H} =- J \sum_{\langle xy \rangle} \vec{s}_x \cdot \vec{s}_y-
D_0\sum_x(\vec{u}_x \cdot \vec{s}_x)^2 + 
K \sum_{x} \sum_a s_{x,a}^4,
\label{HRAM2}
\end{equation}
and the corresponding $\varphi^4$ Hamiltonian
\begin{eqnarray}
{\cal H} = \int d^dx \left[
\frac{1}{2} (\partial_\mu \vec{\varphi})^2 + \frac{1}{2} r \vec{\varphi}^{\,2} 
- D ( \vec{u}\cdot \vec{\varphi})^2 +\frac{1}{4!} v_0 (\vec{\varphi}^{\,2})^2
  + \frac{1}{4!} y_0 \sum_a \varphi_a^4
\right].
\label{heff2} 
\end{eqnarray}
The Hamiltonian (\ref{LGWgen}) was originally 
introduced in Ref.~\cite{MG-82} to describe magnetic systems with 
single-ion anisotropy and nonmagnetic impurities.

There are two interesting particular cases.
First, one may consider an $O(M)$-invariant pure system coupled to an 
isotropic distribution $P(u)$. In this case $K = 0$ in 
Eq.~(\ref{HRAM2})---therefore, $y_0 = 0$---and
$B = 0$ in Eq.~(\ref{mabcd}), so that 
$w_0 = 0$. These conditions are preserved under renormalization by the presence
of the $O(M)$ invariance. Note that this is not the case if $K\not=0$,
i.e. if $y_0\not=0$. Distributions such that $B=0$ 
(these distributions are not necessarily isotropic) give apparently
$w_0 = 0$; however, such a condition is not preserved under renormalization
if $z_0 \not=0$.

A second interesting case corresponds to distributions $P(u)$
such that $A=0$ in Eq.~(\ref{mabcd}). It is easy to show that 
distributions $P(u)$ with this property are simple generalizations of the 
distribution (\ref{Cdistr}). Explicitly, they have the form
\begin{equation}
P(u) = {1\over M} \sum_a f(u_a) \prod_{b\not=a} \delta(u_b), 
\label{RCAM-distr}
\end{equation}
where $f(x)$ is a normalized probability distribution with unit variance. 
If $A = 0$, Eq.~(\ref{hrepl3}) implies $z_0 = 0$. Such a condition
is stable under renormalization. Indeed, the transformation 
$\phi_{ai} \to -\phi_{ai}$ for fixed $a$ and $i$ is a symmetry of the 
Hamiltonian with $z_0 = 0$, but not of the term proportional to $z_0$. 
This symmetry is due to the fact that, for distributions of 
type (\ref{RCAM-distr}), we can write $(\vec{s}\cdot \vec{u})^2 = \sum_a 
u^2_a s_a^2$, which is symmetric under the transformations 
$s_a \to - s_a$ at fixed $u$. In other words, the theory with $z_0 = 0$ 
describes models in which the reflection symmetry of the spins is also 
preserved at fixed disorder. 

In the case of discrete cubic-symmetric distributions of type
(\ref{RCAM-distr}), we have
\begin{equation}
u_0 = {12 D^2\over M^2} ,\qquad w_0 = - {12 B D^2} \; .
\label{RCAMpar} 
\end{equation}
Apparently, these conditions imply 
\begin{equation} 
u_0 > 0, \qquad w_0 < 0,  \qquad M u_0 + w_0 \le 0,
\label{inequalities}
\end{equation}
where the last condition follows from the bound $B \ge 1/M$. The 
equality $M u_0 + w_0 = 0$ is obtained by using distribution (\ref{Cdistr}). 
Relations (\ref{RCAMpar}) and (\ref{inequalities}) 
should be considered as indicative, since
the mapping between the $\varphi^4$ Hamiltonian (\ref{heff2}) and 
the general Hamiltonian (\ref{LGWgen}) gives also rise to higher-order terms. 

It is also interesting to consider the effective continuum Hamiltonian 
corresponding to Eq.~(\ref{HRCAM-gen}). In this case we obtain
\begin{eqnarray}
{\cal H}_{\varphi^4} = \int d^dx \left[
\frac{1}{2} (\partial_\mu \vec{\varphi})^2 + \frac{1}{2} r \vec{\varphi}^{\,2} 
- D \sum_a {q}_a \varphi^2_a +\frac{1}{4!} v_0 (\vec{\varphi}^{\,2})^2 + 
  \frac{1}{4!} y_0 \sum_a \varphi_a^4
\right].
\label{heffRCAM} 
\end{eqnarray}
If $P(q)$ is invariant under the interchanges $q_a \leftrightarrow q_b$, 
we can write for the first moments $M_a = a$ and 
$M_{ab} = b + c \delta_{ab}$. A simple calculation gives again 
the general Hamiltonian (\ref{LGWgen}) with $z_0 = 0$ and 
\begin{equation}
u_0 = {D^2\over2} (a^2 - b), \qquad w_0 = - {c D^2\over2}.
\end{equation}
Since $b+c \ge a^2$, we obtain 
\begin{equation}
u_0 + w_0 \le 0.
\label{rel-u0w0}
\end{equation}
Equality is obtained for $P(q) = \prod_a \delta(q_a - 1)$ 
(in this case however $w_0=0$).
Finally, note that if $c=0$ then we have $w_0=0$.  Such a condition
is stable under renormalization, and thus this class of models is expected
to have a different critical behavior. It corresponds to the one of the 
randomly dilute cubic models discussed in Ref.~\cite{CPV-03-racu}.
Distributions with this property are however quite peculiar. 
They have the general form 
\begin{equation}
P(q) = f(q_1) \prod_{a=2}^M \delta(q_1 - q_a).
\end{equation}

The stability region of the quartic potential in the
$\varphi^4$  Hamiltonian (\ref{LGWCRAM}) is given by the conditions
\begin{eqnarray}
&& N u_0 +   v_0 + N w_0 + y_0 > 0, 
\label{stab1} \\
&& N M u_0 + M  v_0 + N w_0 + y_0 > 0,
\label{stab2} \\
&& u_0 + v_0 + w_0 + y_0 > 0, 
\\
&& M u_0 +M  v_0 + w_0 + y_0 > 0.
\end{eqnarray}
However, as discussed in Ref.~\cite{MG-82}, in the zero-replica limit 
$N\rightarrow 0$, the only relevant stability conditions are those obtained by 
using replica-symmetric configurations. Therefore, for the RCAM one should only 
consider Eqs.~(\ref{stab1}) and (\ref{stab2}) with $N=0$, i.e. 
\begin{eqnarray}
 v_0 + y_0 > 0 && \qquad \hbox{\rm if  } v_0 > 0, \nonumber \\
 M v_0 + y_0 > 0 && \qquad \hbox{\rm if  } v_0 < 0. 
\label{stability-bound}
\end{eqnarray}
Equivalently, the relevant stability conditions can be obtained by 
considering the Hamiltonians (\ref{heff2}) and/or 
(\ref{heffRCAM}).

\section{General renormalization-group properties}
\label{sec2}

\subsection{Fixed points of the theory}
\label{sec2b}

The properties of the RG flow are essentially determined
by its FP's.
Most of them can be identified by considering
the theories obtained  when some of the quartic parameters vanish.
For example, we can easily recognize: 

(a) the O($M\times N$) theory for $v_0=w_0=y_0=0$;

(b) $N$ decoupled O($M$) theories for $u_0=w_0=y_0=0$;

(c) $M$ decoupled O($N$) theories for $u_0=v_0=y_0=0$;

(d) $M\times N$ decoupled Ising theories for $u_0=v_0=w_0=0$;

(e) the $MN$ model (see, e.g., Refs.~\cite{Aharony-76,PV-00}) for $w_0=y_0=0$;

(f) the $NM$ model for $v_0=y_0=0$;

(h) $N$ decoupled $M$-component cubic models for $u_0=w_0=0$;

(i) $M$ decoupled $N$-component cubic models for $u_0=v_0=0$;

(j) $M\times N$-component cubic model for $v_0=w_0=0$;

(k) the randomly dilute $M$-component cubic model (see 
Ref.~\cite{CPV-03-racu}) for $w_0=0$ and $N=0$;

(l) the tetragonal model \cite{Aharony-76,PV-r} for $M=2$ and $w_0=0$.

The FP's of these theories are also FP's 
of the enlarged model (\ref{LGWCRAM}).
Of course, there may also be FP's that are not related to
the above particular cases. Their
presence can be investigated by low-order $\epsilon$ expansion 
calculations.
First-order $\epsilon$-expansion calculations \cite{Aharony-75}  
show the presence of 14 FP's for $M\neq2$ and of 13 FP's for $M=2$.
As in the REIM case,
at two-loop order other $O(\sqrt{\epsilon})$ FP's appear \cite{MG-82}:
4 FP's for $M\neq 2$ and 6 FP's for $M=2$ \cite{DFH-01}.
The two-loop $\epsilon$-expansion results 
are summarized  in Refs.~\cite{DFH-01,HBDFFY-03}.
In Table~\ref{epstable} we report the leading $\epsilon$-expansion
terms for the location of the FP's (in
the minimal-subtraction renormalization scheme)
and the corresponding stability eigenvalues.
The only stable FP's are the O(0) and the REIM FP's, 
which are already present in models (a) and (i), respectively.
These results have also been supported by two-loop fixed-dimension
calculations \cite{DFH-01}.
In order to understand the relevance of the various FP's for the RCAM, 
we need to check which one is accessible from the region
of the quartic parameters relevant for the three-dimensional RCAM.
This issue will be investigated in Sec.~\ref{sec3}
computing and analyzing five-loop series in the framework of the 
fixed-dimension expansion.

\begin{table}[!tbp]
\squeezetable
\caption{Fixed points of the Hamiltonian (\ref{LGWCRAM}) near four dimensions.
    We report the leading nontrivial contribution of the expansion in powers 
    of $\epsilon$, taken from Refs.~\protect\cite{Aharony-75,DFH-01}. Here,
    $K_d=(4\pi)^d \Gamma(d/2)/2$,
    {$\alpha_\pm=(M-4\pm\protect\sqrt{M^2+48})/8$},
    {$\beta_\pm=-(M+12\pm\protect\sqrt{M^2+48})/6$},
    {$A_{\pm\pm}=6\alpha_\pm+3\beta_\pm+M+6$}. 
    The general expressions for the stability eigenvalues of FP's XI-XIV 
    are rather cumbersome. We only report their numerical value 
    for $M=3$.}
\label{epstable} 
{\begin{tabular}{lcccccl}
&& $v^*/K_d$ & $u^*/K_d$ & $w^*/K_d$ & $y^*/K_d$& Stability eigenvalues\\ 
\hline \hline
I&Gaussian & 0 & 0 & 0 & 0& $\o_u=\o_v=\o_w=\o_y=-\e$\\ 
II& O(M) & $\frac{6}{M+8}\epsilon$&   0 & 0 & 0&
$\o_v=\e,\o_u =-\case{4-M}{M+8}\e,\o_w=-\case{4+M}{M+8}\e, 
\o_y =\case{4-M}{M+8}\e$\\ 
III& O(0)  & 0&$\frac{3}{4}\epsilon$ & 0 & 0 &
$\o_u=\e,\o_v=\o_w=\o_y=\e/2$\\ 
IV& O(0)& 0 & 0 & $\frac{3}{4}\epsilon$ & 0&
$\o_u=\o_v=-\e/2,\o_w=\e,\o_y=\e/2$\\ 
V &Ising & 0 & 0 & 0 &$\frac{2}{3}\epsilon$&
$\o_u=\o_v=\o_w=-\e/3,\o_y=\e$\\ 
VI& &$\frac{3}{2(M-1)}\epsilon$ 
&$\frac{3(M-4)}{8(M-1)}\epsilon$&0 &0&
$\o_1=\e,\o_2=\o_y= \case{4-M}{4(M-1)}\e,\o_w=-\case{4+M}{4(M-1)}\e $\\
VII& &0 &$\frac{3}{2}\epsilon$  &
$-\frac{3}{2}\epsilon$& 0&
$\o_1=\o_v=\e,\o_3=\o_y=-\e$\\ 
VIII&Cubic & $\frac{2}{M}\epsilon$&
0& 0 &$\frac{2(M-4)}{3M}\epsilon$&
$\o_u=\o_2= -\case{4-M}{3M}\e,\o_w=-\case{4+M}{3M}\e,\o_4=\e$\\ 
IX& $M\neq 2$&
$\frac{1}{M-2}\epsilon$& $\frac{M-4}{4(M-2)}\epsilon$  & 0 &
$\frac{M-4}{3(M-2)}\epsilon$&
$\o_1=\e,\o_2=\case{4-M}{6(M-2)}\e, \o_w=\case{-(4+M)}{6(M-2)},\o_4= \case{-(4-M)}{6(M-2)}\e$\\ 
X && 0& $\frac{1}{2}\epsilon$
& $-\frac{1}{2}\epsilon$ & $\frac{2}{3}\epsilon$&
$\o_1=\e,\o_v=\o_3=\e/3,\o_4=-\e/3$\\
XI&& $\frac{3}{A_{++}}\epsilon$& 
$\frac{3\alpha_+}{A_{++}}\epsilon$  &
$\frac{3(M+4)}{4A_{++}}\epsilon$ &
$\frac{3\beta_+}{A_{++}}\epsilon$&
$\o_1=\e,\o_2=1.33 \e,\o_3=\o_4 =-1.43 \e$ (for $M=3$)\\
XII& &
$\frac{3}{A_{+-}}\epsilon$&
$\frac{3\alpha_+}{A_{+-}}\epsilon$  &
$\frac{3(M+4)}{4A_{+-}}\epsilon$ &
$\frac{3\beta_-}{A_{+-}}\epsilon$
&$\o_1=\e,\o_2=-\o_3=0.371 \e,\o_4 =-0.344\e$ (for $M=3$)\\ 
XIII&&
$\frac{3}{A_{-+}}\epsilon$&
$\frac{3\alpha_-}{A_{-+}}\epsilon$  &
$\frac{3(M+4)}{4A_{-+}}\epsilon$ &
$\frac{3\beta_+}{A_{-+}}\epsilon $&
$\o_1=\e,\o_2=-\o_3=0.435,\o_4=-0.403\e$ (for $M=3$)\\ 
XIV&&
$\frac{3}{A_{--}}\epsilon$&
$\frac{3\alpha_-}{A_{--}}\epsilon$  &
$\frac{3(M+4)}{4A_{--}}\epsilon$ &
$\frac{3\beta_-}{A_{--}}\epsilon$
&$\o_1=\e,\o_2=\o_3=-3.32\e,\o_4=-3.08\e$ (for $M=3$)\\
XV&REIM& 0 & 0 & $\mp\sqrt{\frac{54}{53}}\sqrt{\epsilon}$ &
$\pm\frac{4}{3}\sqrt{\frac{54}{53}}\sqrt{\epsilon}$ &
$\o_u=\o_v=\o_1=\pm \sqrt{\frac{24}{53}}\sqrt\e,\o_2= 2 \e $ \\ 
XVI&REIM
&0&$\mp\sqrt{\frac{54}{53}}\sqrt{\epsilon}$&  0 &
$\pm\frac{4}{3}\sqrt{\frac{54}{53}}\sqrt{\epsilon}$&
$\o_w=\o_v=-\o_1=\mp \sqrt{\frac{24}{53}} \sqrt\e,\o_2=2\e$\\
XVII& {$M=2$}  & $\mp2\sqrt{\frac{54}{53}}\sqrt{\epsilon}$
& $\pm\sqrt{\frac{54}{53}}\sqrt{\epsilon}$& 0 &
$\pm\frac{4}{3}\sqrt{\frac{54}{53}}\sqrt{\epsilon}$ &
$\o_w=\o_3=-\o_1=\mp \sqrt{\frac{24}{53}} \sqrt\e,\o_2=2\e$
\\
\end{tabular}}
\end{table}

\subsection{Crossover behavior close to the pure spin model}
\label{sec2c}

The O($M$)-symmetric FP located in the
$v$-axis describes the critical properties of the pure spin
system in the absence of cubic anisotropy. 
It is interesting to compute the crossover exponent in the presence 
of random anisotropy.
Setting $t_p\equiv (T-T_p)/T_p$, where $T_p\equiv T_c(D_0=0)$ 
is the critical temperature in the absence of  
anisotropy, in the limit $t_p \to 0$ and $D_0\to 0$ 
the singular part of the free energy can be written as 
\begin{equation}
{\cal F}= |u_t|^{2-\alpha}f(D_0^2 |u_t|^{-\phi_D}),
\label{freeen}
\end{equation}
where $u_t\approx t_p+a_1 D_0^2+a_2 t_p^2+\dots$ is 
the scaling field associated 
with temperature, $\alpha$ is the specific-heat exponent
in the O($M$) theory, $\phi_D$ is the  crossover exponent,
and $f(x)$ is a scaling function.
As a consequence of Eq.~(\ref{freeen}), 
for sufficiently small $D_0$ the critical-temperature shift is given by
\begin{equation}
\Delta T_c(D_0) \equiv 
T_c(D_0)- T_c(0) \approx a D_0^{2/\phi_D} +  b D_0^2 + c D_0^4 + \ldots
\label{tshift}
\end{equation}
The crossover exponent $\phi_D$ is related to the largest
positive RG dimension of the perturbations at the O($M$) FP
that are present in the Hamiltonian (\ref{LGWCRAM}),
i.e. of the terms proportional to $u_0$, $w_0$, and $y_0$. 
For $u_0 = w_0 = y_0 = 0$, the Hamiltonian (\ref{LGWCRAM}) describes 
$N$ decoupled O($M$)-symmetric systems.
The RG dimension of the terms proportional to $u_0$ and $w_0$ 
can be determined by writing \cite{Aharony-75} 
\begin{equation}
\sum_{abij} (u_0 + w_0\delta_{ab}) \phi^2_{ai} \phi^2_{bj} = 
   M (M u_0 + w_0) \sum_{ij} {\cal E}_i {\cal E}_j + 
   w_0 \sum_{ija} {\cal T}_{aai} {\cal T}_{aaj},
\label{dec-u0w0}
\end{equation}
where 
\begin{eqnarray}
{\cal E}_i & \equiv & \frac{1}{M} \sum_a \phi_{ai}^2, 
\nonumber  \\
{\cal T}_{abi} & \equiv & \phi_{ai} \phi_{bi} - \delta_{ab}{\cal E}_i.
\label{def-E-T}
\end{eqnarray}
The bilinears ${\cal E}_i$ and ${\cal T}_{abi}$ 
are respectively the energies and the quadratic spin-2 operators of the 
$N$ decoupled models. If $y_E = 1/\nu$ and $y_T$ are the corresponding 
RG dimensions, the two terms given above have RG dimensions 
$y_u = 2 y_E - 3 = \alpha/\nu$ and $y_w = 2 y_T - 3$. The perturbation 
proportional to $y_0$ does not couple the different replicas and therefore
its RG dimension  $y_y$ is simply the RG dimension of the cubic perturbation,
which is related to the RG dimension of the spin-4 perturbation 
of the $O(M)$ FP \cite{CPV-03,CPV-00}.
Therefore, the $O(M)$ FP is perturbed by three 
terms of RG dimensions $y_u$, $y_w$, and $y_y$, which  
can be determined using known results for the RG dimensions of generic 
operators in an O($M$) theory,
see, e.g., Refs.~\cite{CPV-03-nova,PV-r} for reviews of results.
Since $\alpha$ is negative for $M\geq 2$, we have $y_u < 0$ and therefore
the corresponding term is always irrelevant.
The exponent $y_T$ has been obtained by using field-theoretical \cite{CPV-03}
and Monte Carlo methods \cite{BFMM-96}: field-theoretical analyses give
$y_T=1.766(6)$ for $M=2$ and $y_T=1.790(3)$ for $M=3$, 
while Monte Carlo simulations give $y_T=1.756(2)$ for $M=2$ and 
$y_T=1.787(2)$ for $M=3$.  Correspondingly,
we find $y_w=0.532(12)$ and $y_w = 0.511(6)$ for $M=2$, and 
$y_w=0.580(6)$ and $y_w = 0.573(3)$ for $M=3$.
Therefore, the perturbation proportional to $w_0$ is always relevant.
Finally, using the results of Refs.\cite{CPV-03,CPV-00} for the spin-4
perturbations at the O($M$) FP, we have
$y_y=-0.103(8)$ for $M=2$ and $y_y=0.013(6)$ for $M=3$.
This implies that the $y$-term is irrelevant for $M=2$,
but relevant for $M=3$.
In conclusion, for both $M=2$ and $M=3$, the most relevant quartic perturbation 
is given by the $w$-term, which determines the crossover 
from the pure critical behavior in the limit of small anisotropy strength.
Therefore, $\phi_D= y_w \nu = 0.357(3)$ for $M=2$
and $\phi_D= y_w \nu = 0.412(3)$ for $M=3$. 
In the crossover limit in which $D_0^2 |u_t|^{-\phi_D}$ is held fixed,
the operators with RG dimensions $y_u$ and $y_y$ give rise to scaling 
corrections. In particular, there are corrections proportional to 
$t^{\Delta_y}$, with $\Delta_y \equiv y_y \nu - \phi_D$, 
$\Delta_y = 0.426(6)$ for $M=2$ and $\Delta_y = 0.403(5)$ for $M=3$, 
which are more important than the usual $O(M)$-invariant corrections, 
which vanish as $t^\Delta$, with \cite{PV-r} 
$\Delta \approx 0.54$ for $M=2$ and $\Delta \approx 0.56$ for $M=3$.  

It is worth mentioning that the scaling behavior (\ref{freeen})
with the same crossover exponent $\phi_D$ also holds for 
a RAM with generic distribution $P(\vec{u})$, and in 
particular for the isotropic case.
Indeed, the additional term proportional to $z_0$ appearing 
in the Hamiltonian (\ref{LGWgen})
has the same RG dimension of the $w$-term at the O($M$) FP.
This can be inferred by rewriting 
\begin{equation}
\sum_{abij} \phi_{ai} \phi_{bi} \phi_{aj} \phi_{bj} = 
\sum_{abij} {\cal T}_{abi} {\cal T}_{abj} + M \sum_{ij} {\cal E}_i {\cal E}_j,
\label{dec-z0}
\end{equation}
where ${\cal T}_{abi}$ and ${\cal E}_i$ are defined in 
Eq.~(\ref{def-E-T}). The first term
is the most relevant one and therefore, we obtain $y_z = 2 y_T - 3$ and 
also $y_z=y_w$.

Let us note that the relatively small value of $\phi_D$
makes the measurement of $\phi_D$ from the critical-temperature
shift for small random anisotropy rather  difficult. Indeed, 
in Eq.~(\ref{tshift}) the term $D_0^{2/\phi_D}$ is suppressed with
respect to the first two analytic terms proportional to $D_0^2$ and $D_0^4$, 
since $2/\phi_D\approx 4.9$ (resp.~$2/\phi_D\approx 5.6$) for $M=3$ 
(resp.~$M=2$).
This explains the results of Ref.~\cite{MAGSRJC-93} 
that measured $T_c$ in crystalline Laves-phase
(Dy$_x$Y$_{1-x}$)Al$_2$ for different values of $x$. 
Since \cite{foot-DyAl2} 
$D_0\to 0$ as $x\to 1$, they were able to measure
$\Delta T_c(D_0)$ for $D_0\to 0$. The experimental results were fitted 
assuming $\Delta T_c(D_0)\sim D_0^{2/\psi}$, obtaining $\psi = 0.80(8)$. 
This result is in substantial agreement with the theoretical prediction 
$\Delta T_c(D_0)\sim D_0^{2}$, but does not provide information 
on the crossover exponent $\phi_D$. 

For $M\geq 3$, pure systems with Hamiltonian (\ref{HRAM2}) do not have 
a critical behavior in the $O(M)$ universality class, see, e.g.,
Refs.~\cite{Aharony-76,CPV-00,CPV-03-racu}.
If the system has [111] as easy direction, its critical behavior belongs to 
a different universality class with reduced cubic symmetry, while 
systems with [100] easy axis are expected to show a first-order transition.
In the latter type of systems, randomness may soften the first-order 
transition. This issue shall be discussed in Sec.~\ref{sec3c}.
On the other hand, we now show that 
in cubic systems with [111] easy axis randomness is 
a relevant perturbation and therefore,
for small randomness, these systems show a crossover behavior
with positive exponent $\phi_D$, cf. Eq.~(\ref{freeen}).
For $u_0=w_0=0$ the Hamiltonian (\ref{LGWCRAM}) reduces
to the one for $N$ decoupled systems with cubic symmetry.
The RG dimensions of the terms proportional to $u_0$ and $w_0$ 
at the cubic FP provide the crossover exponent $\phi_D$.
In order to determine them, we use again Eq.~(\ref{dec-u0w0}).
The RG dimension of ${\cal E}_i$ is $y_E=1/\nu$, where
$\nu$ is the correlation-length exponent, while 
that of ${\cal U}_{ai}\equiv {\cal T}_{aai}$, $y_U$, 
is computed in App.~\ref{quadrop}
by resumming its six-loop perturbative expansion.
Thus, the  RG dimensions of the two terms appearing in the right-hand side 
of Eq.~(\ref{dec-u0w0}), we denote them by $y_u$ and $y_w$,
are given by $y_u=2 y_E - 3={\alpha/\nu}$ and $y_w=2 y_U - 3$, respectively.
Since $\alpha<0$ at the cubic FP for any $M\geq 3$, the first term is
irrelevant. On the other hand, the estimates of $y_U$ reported in 
App.~\ref{quadrop} show that $y_w>0$ for any $M\geq 3$. For example
$y_w=0.549(14)$ for $M=3$, and therefore $\phi_D=0.387(14)$.

Note that in a generic cubic-symmetric RAM, one should also consider 
perturbations proportional to $z_0$. 
We use again Eq.~(\ref{dec-z0}). However, in the presence of cubic 
symmetry ${\cal T}_{abi}$, cf. Eq~(\ref{def-E-T}), is not an irreducible 
tensor. One must consider 
separately ${\cal U}_{ai} \equiv {\cal T}_{aai}$ and 
${\cal T}_{abi}$ with $a\not= b$, that have different RG dimensions 
$y_U$ and $y_T$. Therefore, the term proportional to $z_0$ is the 
sum of three terms of RG dimensions $2 y_E - 3 = \alpha/\nu$, 
$2 y_U - 3$, and $2 y_T - 3$. The last one is the largest, so that 
$y_z=2 y_T -3$.
Using the results of App.~\ref{quadrop} for $M=3$, we find
$y_z=0.600(4)$. The exponent $y_z$ is larger than $y_w$. 
Therefore, a  generic cubic-symmetric RAM shows a different crossover 
behavior with crossover exponent $\phi_D=y_z\nu=0.427(3)$. 

\subsection{Stable fixed points}
\label{sec2d}

The critical behavior in the presence of random anisotropy
should be described by the stable FP of the theory (\ref{LGWCRAM})
which is accessible from the RCAM.
The two-loop $\epsilon$-expansion calculations of Ref.~\cite{DFH-01} 
summarized in Sec.~\ref{sec2b} find two 
stable FP's. One of them is located in the $u$-axis, and 
it is associated with the O(0) or self-avoiding-walk
universality class.
This FP is also stable in three dimensions.
Indeed, the terms proportional to $v_0$, $w_0$, and $y_0$ 
are interactions transforming as the spin-4 representation of the 
O($M\times N$) group. Therefore, they have the same RG dimension which is given 
by $y_{v,w,y}=-0.37(5)$,
obtained in Ref.~\cite{CPV-03} from the analysis of six-loop fixed-dimension 
and five-loop $\epsilon$ series.
It was argued in Ref.~\cite{DFH-01}, on the basis of two-loop calculations,
that the O(0) FP is not accessible from the parameter region relevant for the 
RCAM. This will be confirmed 
by the five-loop analysis of the RG flow presented in Sec.~\ref{sec3}.

For $u_0=v_0=0$ the Hamiltonian (\ref{LGWCRAM}) corresponds to 
a cubic-symmetric model and, for $N\to 0$, 
it is the sum of $M$ independent models that are 
the field-theoretical analog of the REIM.
We will now show that the REIM FP is stable in the
theory (\ref{LGWCRAM}). It is sufficient to show that
the terms  proportional to $u_0$ and $v_0$ are irrelevant.
For this purpose, we rewrite
\begin{equation}
\sum_{ijab} (u_0 + v_0 \delta_{ij})\phi^2_{ai}\phi^2_{bj} = 
  N (N u_0 + v_0) \sum_a {\cal E}_a^2 + 
  v_0 \sum_{ai} {\cal U}_{ai}^2,
\label{decomposizione2}
\end{equation}
where ${\cal E}_a=\frac{1}{N}\sum_i \phi_{ai}^2$ and 
${\cal U}_{ai}=\phi_{ai}^2 - {\cal E}_a$.
For $N\to 0$, ${\cal E}_i$ and ${\cal U}_{ai}$ have the same 
RG dimension (see App.~\ref{quadrop} for the proof), $y_E=y_U=1/\nu_r$, 
where $\nu_r$ is the correlation-length
critical exponent of the REIM universality class. Therefore, the RG dimension
of the perturbation is given by $y_{uv} = 2 y_E - 3 = \alpha_r/\nu_r$,
where $\alpha_r$ is the REIM specific-heat exponent.
Since $\alpha_r$ is negative, see the estimates
reported in Refs.~\cite{PV-r,FHY-01,CMPV-03}, the REIM FP is stable. 
Using the recent Monte Carlo results reported in Ref.~\cite{CMPV-03},
we finally arrive at the estimate $y_{uv}\approx -0.07$.
As we shall see in Sec.~\ref{sec3}, the REIM FP turns out to be accessible
to the RCAM, and no other stable FP exists in the region relevant for the RCAM.
Therefore, the REIM universality class
describes the  critical properties of the RCAM
in the case it undergoes a continuous transition.
Estimates of several universal quantities for the
REIM universality class can be found in 
Refs.~\cite{PV-r,FHY-01,CMPV-03,CDPV-03}.
Note, however, that the critical exponent controlling the leading scaling 
corrections differs from the one for the REIM,
which is \cite{BFMMPR-98,CPPV-03} $\Delta_r\approx 0.25$.
In the RCAM the leading scaling correction is due to the Hamiltonian terms
proportional to $u_0$ and $v_0$. They cause
slowly decaying corrections of order $t^{\Delta}$ with
\begin{equation}
\Delta=-\alpha_r=0.049(9).
\end{equation}
If $P(q) = \prod_a P_a(q_a)$, i.e. the probability distributions of the 
variables $q_a$ are independent, 
the stability of the REIM FP can also be proved by starting directly from 
Eq.~(\ref{heffRCAM}). Indeed, such a Hamiltonian corresponds to $M$ 
random-exchange $\varphi^4$ models coupled by the term proportional to $v_0$. 
Such a term has the form $\sum_{ab} {\cal E}_a {\cal E}_b$, where 
${\cal E}_a = {1\over N} \varphi_a^2$ is the energy of the REIM. Therefore,
this perturbation has RG dimension $2/\nu_r - 3 = \alpha_r/\nu_r$, which 
is negative. Thus, the coupling among the models is irrelevant, and thus 
it does not change the universality class of the system. If $P(q)$ does 
not factorize, the $M$ $\varphi^4$ models are also coupled by disorder. 
The above-reported analysis shows that also this coupling is irrelevant, 
its RG dimension being $\alpha_r/\nu_r < 0$. 

As discussed in Sec.~\ref{sec1},
in the case of a generic random cubic-symmetric distribution
$P(\vec{u})$, the Hamiltonian (\ref{LGWgen}) also contains 
the term proportional to $z_0$.
It is important to note that the REIM FP is unstable with
respect to this perturbation, since its RG dimension $y_z$ is positive at the 
REIM FP. The dimension of this perturbation, $y_z$, can be computed by 
rewriting the term proportional to $z_0$ as 
\begin{equation}
\sum_{abij} \phi_{ai} \phi_{bi}\phi_{aj}\phi_{bj} = 
\sum_{ab}\sum_{i\not=j} {\cal T}_{aij} {\cal T}_{bij} + 
\sum_{ab}\sum_{i} {\cal U}_{ai} {\cal U}_{bi} + 
N \sum_{ab} {\cal E}_{a} {\cal E}_{b},
\end{equation}
where ${\cal T}_{aij}\equiv\phi_{ai} \phi_{aj}$ with $i\not= j$. Therefore,
this perturbation is the sum of three terms that have RG dimensions 
$2 y_T - 3$, $2 y_U - 3$, and $2 y_E - 3$. Using the results 
reported in App.~\ref{quadrop}, one finds that the most relevant term is
the first one, so that 
\begin{equation}
y_z=2 y_T - 3.
\end{equation}
Using the estimate $y_T=2.08(3)$ reported in App.~\ref{quadrop}, 
one obtains  
\begin{equation}
y_z=1.16(6), \quad \phi_z\equiv y_z\nu=0.79(4),
\end{equation}
where $\phi_z$ is the corresponding crossover exponent.

\subsection{Critical behavior for infinitely strong random anisotropy}
\label{sec2e}

In this Section, we wish to investigate the general model (\ref{HRCAM-gen})
in the limit of infinite disorder, showing that, under some mild 
hypotheses for the probability distribution $P(q)$, one has REIM 
critical behavior for $M=2$ and $M=3$ and no transition for $M\ge 4$. 
This analysis further confirms the results of Sec.~\ref{sec2d}. 

We first consider the case $D_0\to+\infty$. We suppose that 
the distribution $P(q)$ is such that there is 
only one direction $k$ such that $q_k = \max_b q_b$ (or at least that this
condition is verified with probability one). This is the case if 
the distribution is continuous and is also true for the distribution 
$P(q)$ derived from (\ref{Cdistr}) (note that $q_a = u^2_a$).
Because of the assumption on $P(q)$, 
for $D_0 \to +\infty$ the spin $\vec{s}$ is constrained 
to lie along the $k$ direction, i.e. $s_k = \pm 1$, $s_a = 0$
for $a \not = k$. Thus, in this limit we can rewrite the Hamiltonian
in the following way. At each site we define $M$ Ising variables 
$\sigma_{x,a}$ and $M$ disorder variables $\rho_{x,a}$. The Ising 
variables assume values $\pm 1$, while the disorder variables 
assume values 0 and 1 with probabilities induced by the distribution of 
$\vec{q}$: 
\begin{eqnarray}
\rho_{x,a} = 1 && \; \; \hbox{if } q_{x,a} > q_{x,b}\;\;  \hbox{for every}
                  \; \; b\not= a \; ,
\nonumber \\
\rho_{x,a} = 0 && \; \; \hbox{otherwise.}
\nonumber
\end{eqnarray}
Then, the average value of a quantity ${\cal O}(s_{x,a})$ is given by
\begin{eqnarray}
\overline {\langle {\cal O} \rangle} = 
   \left[ \langle  {\cal O} (\rho_{x,a} \sigma_{x,a}) \rangle_{\sigma} 
        \right]_\rho, 
\end{eqnarray}
where $\left[\cdot\right]_\rho$ indicates the average over the disorder 
variables $\rho_{x,a}$ and $\langle\cdot\rangle_{\sigma}$ indicates the sample
average with Hamiltonian
\begin{equation}
{\cal H} = - J \sum_a \sum_{\langle xy \rangle}
    \sigma_{x,a} \sigma_{y,a} \rho_{x,a} \rho_{y,a}\; .
\label{HRIM-corr}
\end{equation}
If ${\cal O}$ depends only on a single component of the spins, say 
it depends only on $s_{x,1}$, we can integrate out $\sigma_{x,a}$ and 
$\rho_{x,a}$ for $a\ge 2$. Thus, the Hamiltonian becomes a REIM Hamiltonian
with disorder $\rho_{x,1}$. Now, we use the symmetry of $P(q)$ 
to conclude that the probability that $\rho_a$ is 1 must be independent of 
$a$. Since $\sum_a \rho_a$ is always equal to 1, we obtain that  
that $\rho_{x,1} = 1$ with probability 
$1/M$ and $\rho_{x,1} = 0$ with probability $1 - 1/M$. Therefore, 
we obtain that correlation functions of $s_1$ are exactly equal to the 
correlation functions of the site-diluted Ising model with vacancy 
density $1 - 1/M$. Note that this result is not true for correlation functions
that involve different components of the spins. Indeed, the 
Hamiltonian (\ref{HRIM-corr}) corresponds to $M$ REIM, but they are coupled 
by the disorder variables. Thus, these correlation functions
are not simply obtained by multiplying REIM correlation functions. 
These considerations allow us to 
predict the behavior of the model (\ref{HRCAM-gen}) for $D_0 \to +\infty$.
Since the REIM has a continuous transition for spin density $p > p_c$, 
$p_c = 0.3116081(13)$ on a cubic lattice \cite{BFMMPR-99}, 
we predict that the model
has a continuous transition for $M=2$ and $M=3$ and no transition at all 
for $M\ge 4$. 

Let us now consider the opposite case $D_0\to -\infty$. If the distribution 
$P(q)$ is such that there is only one direction $k$ such that 
$q_k = \min_b q_b$, the previous argument applies with no changes. 
Note that distribution (\ref{Cdistr}) does not satisfy this condition
for $M\ge 3$. Indeed, in the limit $D_0\to -\infty$ the spins are constrained
to be orthogonal to $\vec{q}$, and therefore cannot be considered as 
Ising variables. In this particular case, 
the behavior at the transition, if it exists, is not predicted by this argument.

\section{Renormalization-group flow in the quartic-coupling space}
\label{sec3}

\subsection{The fixed-dimension five-loop expansion}
\label{sec3a}

In this section we study the RG flow of the theory (\ref{LGWCRAM}), 
determining the stable FP's and their attraction domain. For this purpose, 
we determine the five-loop perturbative expansion of the 
$\beta$ functions in terms of appropriately defined zero-momentum
quartic couplings at fixed dimension.
In the present case we define $u$, $v$, $w$, and $y$ by writing
\begin{eqnarray}
\Gamma^{(4)}_{aibjckdl}(0) = m
Z_\phi^{-2}  {16 \pi\over 3} \left( u\,R_{MN} A_{aibjckdl} + 
v \,R_M B_{aibjckdl}  + w \,R_N C_{aibjckdl} 
+ y \, D_{aibjckdl} \right),
\label{ren2}  
\end{eqnarray}
where $R_K = 9/(8+K)$, $A$, $B$, $C$, and $D$ are appropriate tensors
defined so that at tree level, $u_0 = m u\,R_{MN}$, $v_0 = m v \,R_M$, 
$w_0 = m w \,R_N$, and $y_0 = m y$.
The mass $m$ and the renormalization constant $Z_\phi$ are 
defined by
\begin{equation}
\Gamma^{(2)}_{aibj}(p) =
  \delta_{ab} \delta_{ij} Z_\phi^{-1} \left[ m^2+p^2+O(p^4)\right].
\label{ren1}  
\end{equation}
Here $\Gamma^{(4)}$ and $\Gamma^{(2)}$ are 
the four- and two-point one-particle irreducible correlation functions.

We computed the $\beta$-functions to five loops, 
which required the calculation of 161 Feynman diagrams.
We employed a symbolic manipulation program, which  generated the diagrams 
and computed the symmetry and group factors of each of them.
We used the numerical results compiled in Ref.~\cite{NMB-77}
for the integrals associated with each diagram.
We do not report the series for  generic values of $N$ and $M$,
but only for the the physically interesting cases $N=0$ and $M=3,2$. 
The series for generic $N$ and $M$ are available on request. 
The coefficients of the five-loop series of the
$\beta$-functions $\beta_u$, $\beta_v$, $\beta_w$, and 
$\beta_y$ are reported in Tables \ref{tabMC3u}-\ref{tabMC3y} for $M=3$ 
and Tables \ref{tabMC2u}-\ref{tabMC2y} for $M=2$. 
We performed the following checks:
\begin{itemize}
\item[(i)]
The series are symmetric under the transformation $v\leftrightarrow w$
and $M\leftrightarrow N$;
\item[(ii)]
For $w=y=0$  and for $v=y=0$ (with $N\leftrightarrow M$),
we recover the series for the $MN$ model reported to six loops in 
Ref.~\cite{PV-00}; 
\item[(iii)]
For $v=w=0$ we obtain 
the series for the ($M\times N$)-component cubic model,
reported to six loops in Ref.~\cite{CPV-00};
\item[(iv)]
We obtain the series for the randomly dilute cubic model 
for $w=0$ and $N=0$ and those of the tetragonal model
for $w=0$ and $M=2$. These series are reported to six loops in 
Refs.~\cite{CPV-03-racu,PV-r}.
\item[(v)] 
For $N=0$ the series satisfy the identities
\begin{eqnarray}
&& \beta_u(u,0,w,y) + \beta_w(u,0,w,y) = \beta_{{\rm REIM},u}(u+w,y), 
\nonumber \\
&& \beta_y(u,0,w,y) = \beta_{{\rm REIM},y}(u+w,y),
\label{identities}
\end{eqnarray}
where $\beta_{{\rm REIM},u}(u,y)$ and $\beta_{{\rm REIM},y}(u,y)$ are the 
$\beta$-functions of the REIM model obtained by setting $v=w=0$ and $M=1$.
The corresponding six-loop series are reported in Ref.~\cite{CPV-00}.
These identities are proved in App.~\ref{betaided}.
\item[(vi)]
At two loops the series agree with the expansions reported in 
Ref.~\cite{DFH-01}. 
\end{itemize}

Perturbative series are divergent and thus a careful analysis is needed in
order to obtain quantitative predictions. In the case of systems without
randomness, they are conjectured to be Borel summable and this allows one to 
use the Pad\'e-Borel method or methods based on a conformal mapping 
\cite{LGZ-80}. In random systems, the perturbative approach faces additional
difficulties: the perturbative series are expected not to be 
Borel summable \cite{BMMRY-87,AMR-99}. Nonetheless, in the REIM case 
quite reasonable estimates of the critical exponents have been 
obtained by using the fixed-dimension expansion in $d=3$ (see, e.g., 
Refs.~\cite{PV-r,FHY-01}). Similarly, 
the usual resummation methods applied to the RCAM expansions 
give quite stable results, at least when the quartic couplings are not too 
large, giving us confidence on the correctness of the conclusions.

\subsection{The RG trajectories}
\label{sec3b}

The knowledge of the $\beta$ functions allows us to study the RG flow
in the space of the quartic renormalized couplings $u$, $v$, $w$, and $y$.
For this purpose we follow closely Ref.~\cite{CPPV-03}.
The RG trajectories are lines starting from the Gaussian FP
(located at $u=v=w=y=0$) along which the quartic Hamiltonian parameters
$u_0$, $v_0$, $w_0$, and $y_0$ are kept fixed.
The RG curves in the coupling space depend on 
three independent ratios of the quartic couplings. 
The RG trajectories can be determined by
solving the differential equations
\begin{eqnarray}
-\lambda {d g_i\over d\lambda} = \beta_{g_i},
\label{rgflow}
\end{eqnarray}
where $g_i=u,v,w,y$, and
$\lambda\in (0,\infty)$, with the initial conditions
\begin{eqnarray}
g_i(0) = 0 ,\qquad
\left. {d g_i\over d\lambda} \right|_{\lambda=0} = s_i,
\end{eqnarray}
where $s_1=s_u\equiv u_0/|v_0|$, 
$s_3=s_w\equiv w_0/|v_0|$, $s_4=s_y\equiv y_0/|v_0|$, and
$s_2 = +1$ if $v_0 > 0$, $s_2 = -1$ if $v_0 < 0$.
The functions $g_i(\lambda,s_i)$ provide the RG trajectories in
the renormalized-coupling space.
The attraction  domain of a FP $g_i^*$
is given by the values of $u_0$, $v_0$, $w_0$, and $y_0$ corresponding to 
trajectories ending at $g_i^*$, i.e. trajectories for which 
$g_i(\lambda=\infty,s_i) = g_i^*$.
We recall that 
the O($M$) FP is located in the $v$-axis
at $v^*\approx 1.40,1.39$ for $M=2,3$ respectively \cite{GZ-98,PV-98};
the O(0) FP lies in the $u$-axis at $u^*\approx 1.39$ 
(Refs.~\cite{GZ-98,PV-98});
the REIM FP lies in the $w$-$y$ plane at $w^*\approx -0.7$ and 
$y^*\approx 2.3$ (we report here the field-theoretical estimates of 
Ref.~\cite{PV-00}; Monte Carlo estimates are given in Ref.~\cite{CMPV-03}).

\subsection{Results}
\label{sec3c}

\begin{figure*}[t!]
\includegraphics[width=15truecm]{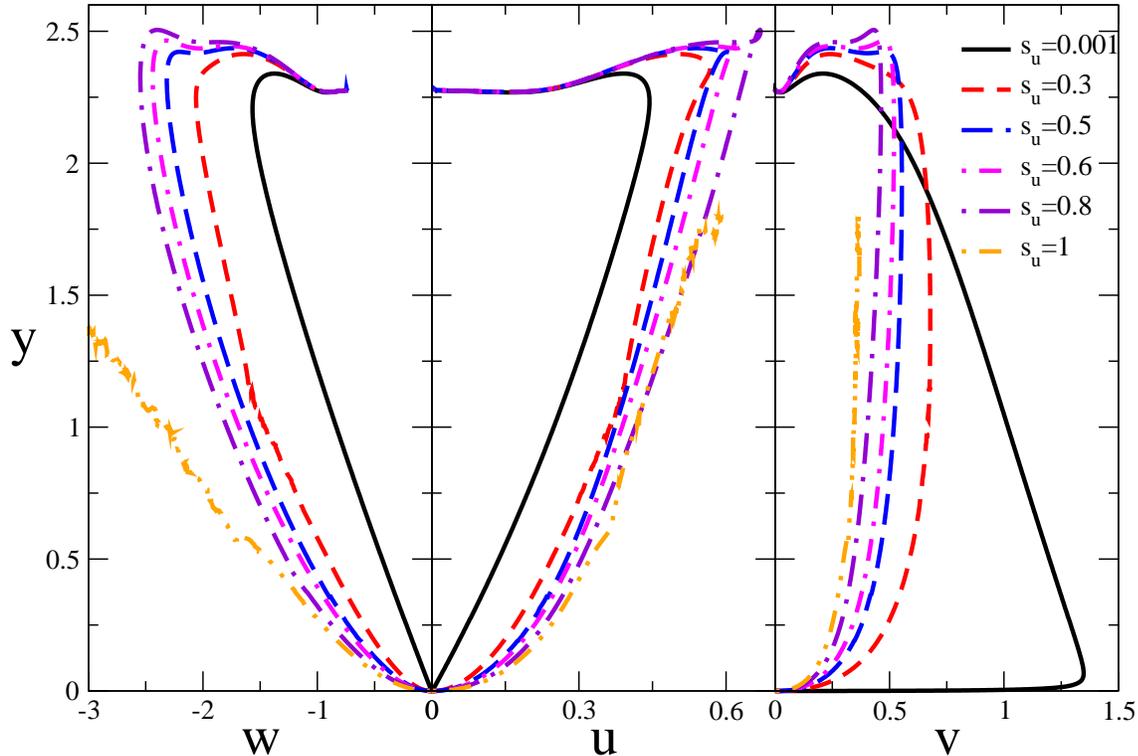}
\caption{
Projections of the RG flow for the three-component case, $M=3$,
in the $y$-$w$, $y$-$v$ and $y$-$u$ planes, as a function
of $s_u \equiv u_0/v_0$, for $u_0 > 0$, $v_0 > 0$, and $w_0 < 0$.
Here $y_0=0$ and $3 u_0 + w_0 =0$.
The REIM FP corresponds to $u^* = v^* = 0$, 
$w^*\approx -0.7$, and $y^*\approx 2.3$ (Ref.~\protect\cite{PV-00}).
}
\label{cross1}
\end{figure*}

In this Section we report our analyses of the five-loop perturbative series.
We have resummed the $\beta$-functions by using the Pad\'e-Borel method.
The major numerical problem we faced was the fact that most of the 
approximants were defective in some region of the coupling space,
forbidding a complete study of the RG flow. 
This is not unexpected since the perturbative series are not Borel summable.
Approximant [3/1] for $\beta_u$, [4/1] for $\beta_v$, and [3/2] for 
$\beta_w$ with $b=1$ \cite{def-PB} were not defective 
in all the region of the RG flow we considered
(in some cases the Pad\'e
approximant to the Borel transform had a pole on the positive real axis 
but far from the origin, in a region that gives a negligible contribution to
the resummed function). 
On the other hand, all approximants for $\beta_y$ 
were defective somewhere in the region we wished to investigate. 
For $\beta_y$ we used approximant [3/2] with $b = 1$, that had the 
least extended defective region. All results we present here were
obtained by using these approximants. It must be stressed that other choices
gave results that were similar in the regions in which they were well-defined.

First, we checked the general results reported in Sec.~\ref{sec2}.
We considered the O($M$), cubic, O(0), and REIM FP's and 
for each of them we determined the stability eigenvalues. 
The results are in full agreement with the conclusions of 
Sec.~\ref{sec2}, confirming that the O(0) and the REIM FP's 
are stable. Then, we looked for additional FP's beside those identified by 
the $\epsilon$-expansion analysis of Sec.~\ref{sec2b}.
For this purpose we considered the RG flow starting from arbitrary values of 
$u,v,w,y$. We only observed runaways trajectories or 
a flow towards either the REIM or the O($0$) FP's, confirming that 
the REIM and the O($0$) are the only stable FP's.
In particular, trajectories corresponding to  Hamiltonian parameters
$w_0 < 0$, $u_0 > 0$, 
$u_0 + w_0 < 0$, and that satisfy the stability bound (\ref{stability-bound})
never flow towards the O($0$) FP, which is therefore 
not accessible from this region.
They either flow towards the REIM FP or apparently run away towards infinity.

For the purpose of illustration, we first consider the case
$y_0=0$, $w_0/u_0= -M$, and $v_0 > 0$, which apparently 
corresponds to the model (\ref{heff}) with distribution 
(\ref{Cdistr}), cf. Eq.~(\ref{RCAMpar}) (note that $B=1/M$ in this case).
In Fig. \ref{cross1} we show the RG 
trajectories for $M=3$ for several values of $s_u>0$.
The approximant of $\beta_y$ is defective for $0.05\lesssim s_u\lesssim 0.3$ 
and $y$ close to 1. This explains the sudden change of direction of the 
trajectory with $s_u=0.3$ in Fig. \ref{cross1} when $y$ is close to 1.
For $M=3$ (resp.~$M=2$) the RG trajectories appear to  approach the REIM FP
for $0<s_u\lesssim 0.9$ (resp.~$0<s_u\lesssim 1.4$).
For larger values of $s_u$ the flow runs close to regions in which 
some approximant is defective. Apparently, trajectories flow towards 
infinity, but this could be an artifact of the resummation. 
In any case, if true, this would imply that the corresponding systems do not 
undergo a continuous transition.
As a consequence, since $s_u$ is directly related to the 
anisotropy strength $D$, the continuous transition would be expected to 
disappear for sufficiently large values of $D$. These conclusions do not 
immediately apply to fixed-length spin systems, i.e. to the Hamiltonian 
(\ref{HRAM}) since in this case \cite{foot-v0-inf} 
$v_0 = +\infty$. Thus, it is 
not clear which is the correct value of $s_u$ even for $u_0 = \infty$. 
The critical behavior of this system for strong disorder has been discussed 
in Sec.~\ref{sec2e}. 

The qualitative picture does not change if we do not require $w_0/u_0 = - M$
and $s_y = 0$. For instance, we can consider the case $w_0/u_0= - M$, 
$v_0 > 0$, and arbitrary $s_y$.
We are able to resum reliably the perturbative series for
$s_y> - 0.7$ and there we observe that some trajectories flow towards 
the REIM FP, while others run away to infinity. 
The attraction domain of the REIM FP enlarges with increasing $s_y$:
it is approximarely bounded by 
$s_u \lesssim 0.9+ 0.6s_y$ for $M=3$ 
($s_u \lesssim 1.4 + 1.4s_y$ for $M=2$) in the region 
$-0.6 \lesssim s_y  \lesssim 0.3$ ($-0.7 \lesssim s_y  \lesssim 1$).
For larger value of $s_y$, the attraction domain becomes even larger
and extends beyond the lines reported above.
For $s_y\lesssim -0.6$ some approximants become defective and we cannot
determine reliably the RG flow. As in the case $s_y = 0$, there is 
some evidence that trajectories flow towards infinity for 
$s_y \lesssim - 1$, while for $-1 \lesssim s_y\lesssim -0.6$ they may still
flow to the REIM FP.

Then, we have investigated the behavior for $v_0 < 0$, although this 
region does not appear to be of physical interest. As in the pure case, 
all trajectories apparently run away to infinity. 

Finally, let us discuss whether O($0$) critical behavior can be observed by 
appropriately tuning the model parameters. We have investigated this question 
in detail. We find that the O($0$) FP can be reached only if 
$u_0 + w_0 > 0$, irrespective of the other parameters as long as 
$u_0 > 0$ and $w_0 < 0$. This result can be proved straightforwardly 
in the limiting case $v_0 = 0$. Indeed, since $\beta_v = 0$ for $v = 0$, 
if we start with $v_0 = 0$ the flow will be confined in the 
hyperplane $v = 0$. But for $v = 0$ we can use identities
(\ref{identities}) that show that the flow for the couplings 
$u+w$ and $y$ is identical to the flow observed in the 
random-exchange $\varphi^4$ theory. Therefore, for $u_0 + w_0 = 0$ 
we observe pure Ising behavior, while for $u_0 + w_0 < 0$ 
(resp. $u_0 + w_0 > 0$) RG trajectories flow towards the REIM 
(resp. O($0$)) FP. Therefore, if $w_0/u_0 < -1$, as implied by 
Eq.~(\ref{rel-u0w0}), only REIM critical behavior can be observed.
For $v_0 > 0$, similar conclusions are obtained numerically: The attraction 
domain of the O($0$) FP is included in the region $w_0/u_0 > - c$, where 
the constant $c$ is positive and smaller than 1, depends on $s_w$, and tends 
to 1 as $s_w \to -\infty$, i.e. $v_0 \to 0$. 

In conclusion, our analysis gives a full picture of the critical behavior 
for cubic magnets that have $v_0 > 0$---we have $v_0=+\infty$ for fixed-length
spins \cite{foot-v0-inf}.
For $M=2$ the pure system has a critical XY transition for $s_y > - 2/3$,
an Ising transition for $s_y = -2/3$, and a first-order transition for 
$-1< s_y < -2/3$ [values of the parameters such that $s_y \le -1$ 
are not allowed since they do not satisfy the stability bound 
(\ref{stability-bound})]. 
Note that first-order transitions cannot be 
observed in the pure model (\ref{HRAM2}) with fixed-length spins. 
Indeed, for the strongest possible negative anisotropy, $K = -\infty$, 
the Hamiltonian can be written as two decoupled Ising models \cite{foot-XY}, 
and thus the system with $K=-\infty$ exactly corresponds to $s_y = -2/3$. 
As a consequence, finite values of $K$ have $s_y > - 2/3$, and therefore
the model is expected to have always an XY transition. 
Randomness changes the critical behavior. 
For small randomness and small anisotropy, we always predict REIM critical 
behavior, while for 
large disorder (unless we consider fixed-length spins, cf. Sec.~\ref{sec2e})
we do not expect a continuous transition. 
Note that the behavior of systems with $-1< s_y < -2/3$ remains unclear
since in this region we are not able to resum reliably the perturbative 
expansions. In particular, we cannot clarify if softening occurs.
A Monte Carlo simulation ~\cite{Fisch-93} found that model (\ref{HRAM2})
with $K=-\infty$ has a continuous transition for small disorder, in agreement
with our results.

For $M=3$ we expect a continuous transition for $s_y \ge 0$ and 
a first-order one for $s_y < 0$. If we add randomness to systems with 
$s_y>0$, the continuous transition survives but now belongs to the 
REIM universality class; for large disorder the transition may disappear.
For $s_y < 0$ one may observe softening, i.e. the first-order transition
may be changed into a continuous one by small disorder. 
Note that softening always occurs
for infinite disorder, $D_0 = +\infty$, in the model (\ref{HRCAM-gen}),
independently of the sign of $y_0$, under mild assumptions on the 
distribution $P(q)$, see Sec.~\ref{sec2e}.

\section*{Acknowledgments}

PC acknowlegdes financial support from EPSRC Grant No. GR/R83712/01

\appendix

\section{Renormalization-group dimensions of bilinear operators in the 
cubic-symmetric $\mathbf{\Phi^4}$ theory}
\label{quadrop}

In this appendix we compute the 
RG dimensions of bilinear operators in the cubic-symmetric theory
\begin{equation}
{\cal H}_c = \int d^d x 
\left\{ {1\over 2}(\partial_\mu \varphi(x))^2 + {1\over 2} r \varphi(x)^2 + 
{1\over 4!} u_0 \left[ \varphi(x)^2\right]^2 +
{1\over 4!} v_0 \sum_{i=1}^N \varphi_i(x)^4 \right\},
\label{Hphi4cubic}
\end{equation}
where $\varphi$ is an $N$-component field.
The bilinear operators can be written in terms of tensors belonging to different
irreducible representations of the cubic group:
\begin{eqnarray}
&&E =\frac{1}{N} \sum_k \varphi_k^2 , \\ 
&&U_i = \varphi_i^2 - \frac{1}{N} \sum_k \varphi_k^2, \\
&&T_{ij} = \varphi_i  \varphi_j, \quad i\neq j.
\end{eqnarray}
The RG dimension of the energy operator $E$ is $y_E=1/\nu$, 
where $\nu$ is the correlation-length exponent.
The RG dimensions of the operators
$U_i$ and $T_{ij}$, respectively $y_U$ and $y_T$,
in the cubic-symmetric theory (\ref{Hphi4cubic}) will be computed
below. Note that in  O($N$)-symmetric theories the tensors $U_i$ and $T_{ij}$
belong to the same irreducible representation and therefore $y_T = y_U$. 
In cubic systems this is no longer the case. 

In order to  compute $y_U$ and $y_T$, we consider the perturbative approach
in terms of the zero-momentum quartic couplings $u$ and $v$ at fixed dimension.
We refer the reader to Ref.~\cite{CPV-00} for notations
and definitions; there one can also find the six-loop perturbative expansion
of the $\beta$-functions and of the RG functions associated with the 
standard exponents.
In order to compute the RG dimensions of $U_i$ and $T_{ij}$,
we consider the related RG functions $Z_U$ and $Z_T$, defined in terms of 
the zero-momentum one-particle irreducible two-point functions 
$\Gamma_U^{(2)}(0)$ and 
$\Gamma_T^{(2)}(0)$ with an insertion of the operator $U_i$ and $T_{ij}$, 
respectively, i.e.
\begin{equation}
\Gamma_U^{(2)}(0)_{i,kl} = Z_U^{-1} \; B_{i,kl},\qquad
\Gamma_T^{(2)}(0)_{ij,kl} = Z_T^{-1} \; A_{ij,kl},
\end{equation}
where $B$ and $A$ are appropriate constant tensors such that $Z_U=Z_T=1$ at 
tree level.
Then, we compute  the RG functions $\eta_U$ and $\eta_T$ defined by
\begin{equation}
\eta_{U,T}(u,v) = \left. {\partial \ln Z_{U,T}\over \partial \ln m}\right|_{u_0,v_0}
= \beta_u {\partial \ln Z_{U,T} \over \partial u} +
\beta_v {\partial \ln Z_{U,T} \over \partial v},
\end{equation}
where $\beta_u$ and $\beta_v$ are the $\beta$-functions.

We computed the functions $\Gamma^{(2)}_{U,T}(0)$ to six loops. 
The resulting six-loop series of $\eta_{U,T}(u,v)$ are
\begin{eqnarray}
&&\eta_U(u,v)= -\frac{2 u}{8+N}-\frac{1}{3} v+u^2\frac{12+2 N}{3(8+N)^2}+ 
\frac{4}{3 (N+8)} u v +\frac{2}{27} v^2+ \sum_{ij} e^U_{ij} u^i v^j, 
\label{yus}\\
&&\eta_T(u,v)= -\frac{2 u}{8+N} +u^2\frac{12+2 N}{3(8+N)^2}+ 
\frac{4}{9 (N+8)} u v + u \sum_{ij} e^T_{ij} u^i v^j, 
\label{yts}
\end{eqnarray}
where $u$ and $v$ are normalized so that
\begin{equation}
m u = {8+M\over 48 \pi} Z_u u_0,\qquad
m v = {3\over 16 \pi} Z_v v_0,\qquad Z_{u,v}=1 + O(u,v), 
\end{equation}
and the coefficients $e^{U}_{ij}$ an $e^{T}_{ij}$ 
are reported in Tables~\ref{tu} and \ref{tt} respectively.
Note that $u$ and $v$ correspond to $\bar{u}$ and $\bar{v}$ in Ref.~\cite{CPV-00}.
The RG dimensions  $y_U$ and $y_T$ are obtained by
$y_{U,T} = 2 + \eta_{U,T} - \eta$, where $\eta_{U,T}$ is the
value obtained by resumming the corresponding series, 
evaluating it at
$u=u^*$, $v=v^*$, where ($u^*$,$v^*$) is the stable FP.

In the case  of the REIM, i.e., in the limit $N\rightarrow 0$, one has
$y_U=y_E=1/\nu$. Indeed, 
$\langle \varphi_i^2 \varphi_k \varphi_l\rangle^{1PI}$ and 
$\langle \sum_i \varphi_i^2 \varphi_k \varphi_l\rangle^{1PI}$ 
are both finite and nonvanishing for $N\to 0$. Therefore, we have for $N\to0$
\begin{equation}
\langle N U_i \varphi_k \varphi_l\rangle^{1PI} = 
- \langle \sum_j \varphi_j^2 \varphi_k \varphi_l\rangle^{1PI} + 
  O(N) = - \langle N E \varphi_k \varphi_l\rangle^{1PI} + O(N).
\end{equation}
Using the Monte Carlo estimate $\nu=0.683(3)$ (Ref.~\cite{CMPV-03}),
we obtain  $y_U=1.464(6)$.
The series for $\eta_T$ was already reported in Ref.~\cite{CPV-03-crexp}.
Its analysis provided the estimate $y_T=2.08(3)$.

For $N=2$ the stable FP of the cubic theory is the O(2) FP, so 
$y_U=y_T=1.766(3)$ \cite{CPV-03}. For $N\geq 3$ the O($N$) FP is unstable 
and the RG trajectories flow toward another FP characterized
by a discrete cubic symmetry.
The analysis of the series, using the same procedure reported in
Ref.~\cite{CPV-00}, gives the estimates
\begin{eqnarray}
&&y_U(N=3)= 1.774(7),\qquad y_T(N=3)=1.800(2),\\
&&y_U(N=4)= 1.696(8),\qquad y_T(N=4)=1.874(3).
\end{eqnarray}

\section{Some renormalization-group identities}
\label{betaided}

In this Appendix we prove relations (\ref{identities}).
Morever, we show that, in the limit $N\to 0$, 
the RG functions do not depend on $M$ for $v=0$.

Let us consider the Hamiltonian for $v_0 = 0$ and rewrite 
\begin{eqnarray}
&& \exp\left[-{1\over 4!} \sum_{ij,ab} 
    (u_0 + w_0 \delta_{ab} + y_0 \delta_{ab} \delta_{ij}) \phi^2_{ai} 
    \phi^2_{bj}\right] \sim
   \int d\lambda d\rho_a d\sigma_{ai} 
\nonumber \\
&& \qquad
  \exp\left[{1\over 2}
     (\lambda^2 + \sum_a \rho_a^2 + \sum_{ai}\sigma_{ai}^2) + 
    {1\over 2 \sqrt{3}}\sum_{ai} 
    (\sqrt{u_0} \lambda + \sqrt{w_0}  \rho_a + \sqrt{y_0} \sigma_{ai}) 
     \phi_{ai}^2\right],
\end{eqnarray}
where $a$ (resp. $i$) runs from 1 to $M$ (resp. $N$) and 
$\lambda$, $\rho_a$, and $\sigma_{ai}$ are auxiliary fields.
Then, let us consider the $n$-point irreducible correlation function. 
We will show that it has the form 
\begin{eqnarray}
\langle \phi_{a_1i_1}\ldots \phi_{a_ni_n}\rangle
= \sum_\alpha c_\alpha Q_\alpha^{{a_1i_1}\ldots {a_ni_n}} ,
\end{eqnarray}
where $Q_\alpha$ are group tensors (products of Kronecker 
deltas), the scalar factors $c_{\alpha}$ do not depend on $M$, and the sum 
runs over all possible independent group tensors. In terms of the 
auxiliary fields, Feynman diagrams contain loops of $\phi$-fields and $n/2$ 
open lines of $\phi$-fields connected by the auxiliary-field lines. 
The group factor associated with each diagram is computed as follows. 
One assigns indices $ai$ to $\phi$ and $\sigma$ propagators and 
indices $a$ to $\rho$ propagators, considers the product of 
the factors $V$ (reported below) associated with each
vertex, and sums over all assigned indices. In order to prove the 
$M$-independence we will show that, because of the Kronecker $\delta$'s 
appearing in the diagrams, none of these sums is effectively performed in the 
limit $N\to 0$, so that no factor of $M$ can appear.
The factors $V$ are given by:
\begin{eqnarray}
V(\lambda, \phi_{ai}, \phi_{bj}) = 
   {\sqrt{u_0}\over \sqrt{3}} \delta_{ab} \delta_{ij} , \nonumber \\
V(\rho_c, \phi_{ai}, \phi_{bj}) = 
   {\sqrt{w_0}\over \sqrt{3}} \delta_{abc} \delta_{ij}, \nonumber \\
V(\sigma_{ck}, \phi_{ai}, \phi_{bj}) = 
   {\sqrt{y_0}\over \sqrt{3}} \delta_{abc} \delta_{ijk},
\end{eqnarray}
where $abc$ (resp. $ijk$) run from 1 to $M$ (resp. $N$).
First, note that all $\phi$ loops must contain at least 
a $\sigma\phi\phi$ vertex, otherwise by summing over the indices of the $\phi$
fields appearing in the loop one obtains a factor of $N$. As a consequence, 
all loops give rise to a very simple effective vertex for the 
auxiliary fields:
\begin{eqnarray}
\langle \lambda\ldots\lambda\rho_{a_1}\ldots\rho_{a_n}
      \sigma_{b_1i_1}\ldots\sigma_{b_mi_m}\rangle 
 \sim \delta_{a_1\ldots a_n b_1\ldots b_m}
      \delta_{i_1\ldots i_m},
\end{eqnarray}
where we have only written the dependence on the group indices.
Then, given a diagram, let us consider the reduced diagram in which all 
$\phi$ loops are replaced by the corresponding effective vertices. 
The $\sigma_{ai}$ propagators form several connected paths.
It is easy to convince oneself that each of these paths must end at an open
$\phi$ line, otherwise, by summing over the indices $i$ associated with the 
$\sigma$ lines, one obtains factors of $N$. As a consequence, all effective 
vertices are connected by $\sigma$ propagators to the open $\phi$ lines.
Therefore, by summing over the indices associated with the $\phi$ and 
$\sigma$ propagators one obtains expressions in which all remaining indices 
(those related to $\rho$ propagators) are equal to external ones and are not
summed over.  We have thus proved that correlation functions expressed in 
terms of 
the bare parameters do not depend on $M$. Since $R_{MN}$ is $M$ independent
for $N\to 0$, this result extends trivially to the RG functions expressed 
in terms of the renormalized couplings.

To prove identities (\ref{identities}) we now exploit the $M$ independence.
For $M=1$ the theory corresponds to the REIM model with couplings 
$u_0 + w_0$ and $y_0$. Since $R_{MN} = R_{N}$ for $N\to 0$,
$(u + v)/m$ is a function of $u_0 + w_0$.  The result follows immediately.



\begin{table}[t]
\squeezetable
\caption{ 
The five-loop series of
$\beta_u = m \partial u/\partial m$ for $M=3$.
Setting $\beta_u \equiv \sum_{l=0} \beta_{u,l}$, 
where $l$ is the number of loops, we report
$\beta_{u,l}$ up to $l=5$.
}
\label{tabMC3u}
\begin{tabular}{c|l}
\multicolumn{1}{c}{$l$}&
\multicolumn{1}{c}{$\beta_{u,l}$}\\
\tableline
0&$-u$\\
\hline
1&$ u^2 + \case{10}{11}u\,v + \case{1}{2}u\,w + \case{2}{11}v\,w + 
   \case{2}{3}u\,y$\\
\hline
2&$-\case{95}{216}u^3 - \case{250}{297}u^2\,v 
- \case{460}{3267}u\,v^2 - 
   \case{25}{54}u^2\,w - \case{41}{99}u\,v\,w 
- \case{8}{363}v^2\,w - 
   \case{23}{216}u\,w^2 - \case{1}{33}v\,w^2 
- \case{50}{81}u^2\,y - 
   \case{184}{891}u\,v\,y - 
\case{23}{81}u\,w\,y - \case{92}{729}u\,y^2$\\
\hline
3&$0.389923\,u^4 + 1.16913\,u^3\,v + 0.621933\,u^2\,v^2 + 0.10714\,u\,v^3 + 
   0.64302\,u^3\,w + 1.1669\,u^2\,v\,w + 0.41177\,u\,v^2\,w + 
   0.02168\,v^3\,w $\\&$+ 0.35637\,u^2\,w^2 + 0.394263\,u\,v\,w^2 + 
   0.048431\,v^2\,w^2 + 0.0730773\,u\,w^3 + 0.0254122\,v\,w^3 + 
   0.857364\,u^3\,y + 0.912169\,u^2\,v\,y $\\&$+ 0.235717\,u\,v^2\,y + 
   0.95032\,u^2\,w\,y + 0.597082\,u\,v\,w\,y + 0.023813\,v^2\,w\,y + 
   0.292309\,u\,w^2\,y + 0.0327432\,v\,w^2\,y + 0.46739\,u^2\,y^2 $\\&$+ 
   0.230387\,u\,v\,y^2 + 0.299507\,u\,w\,y^2 + 0.00418781\,v\,w\,y^2 + 
   0.090449\,u\,y^3$\\
\hline
4& $-0.447316\,u^5 - 1.83254\,u^4\,v - 1.7466\,u^3\,v^2 - 0.575817\,u^2\,v^3 - 
   0.076344\,u\,v^4 - 1.0079\,u^4\,w - 2.83639\,u^3\,v\,w - 
   1.9044\,u^2\,v^2\,w $\\&$- 0.380099\,u\,v^3\,w - 0.0145582\,v^4\,w - 
   0.885278\,u^3\,w^2 - 1.74084\,u^2\,v\,w^2 - 0.695073\,u\,v^2\,w^2 - 
   0.038822\,v^3\,w^2 - 0.37419\,u^2\,w^3 $\\&$- 0.451159\,u\,v\,w^3 - 
   0.0547698\,v^2\,w^3 - 0.06557\,u\,w^4 - 0.0232153\,v\,w^4 - 
   1.34386\,u^4\,y - 2.56167\,u^3\,v\,y - 1.2668\,u^2\,v^2\,y $\\&$- 
   0.223943\,u\,v^3\,y - 2.36074\,u^3\,w\,y - 3.0627\,u^2\,v\,w\,y - 
   0.87616\,u\,v^2\,w\,y - 0.023185\,v^3\,w\,y - 1.49676\,u^2\,w^2\,y - 
   1.14351\,u\,v\,w^2\,y $\\&$- 0.0520405\,v^2\,w^2\,y - 0.349706\,u\,w^3\,y - 
   0.0459449\,v\,w^3\,y - 1.2213\,u^3\,y^2 - 1.23262\,u^2\,v\,y^2 - 
   0.32717\,u\,v^2\,y^2 - 1.5635\,u^2\,w\,y^2 $\\&$- 0.888055\,u\,v\,w\,y^2 - 
   0.0135895\,v^2\,w\,y^2 - 0.550107\,u\,w^2\,y^2 - 0.0186856\,v\,w^2\,y^2 - 
   0.476234\,u^2\,y^3 - 0.252712\,u\,v\,y^3 $\\&$- 0.335525\,u\,w\,y^3 - 
   0.00289791\,v\,w\,y^3 - 0.0754467\,u\,y^4$\\ 
\hline
5& $0.633855\,u^6 + 3.32208\,u^5\,v + 4.6464\,u^4\,v^2 + 2.53461\,u^3\,v^3 + 
   0.702909\,u^2\,v^4 + 0.0816358\,u\,v^5 + 1.82714\,u^5\,w + 
   7.04207\,u^4\,v\,w $\\&$+ 7.40684\,u^3\,v^2\,w + 2.97466\,u^2\,v^3\,w + 
   0.499013\,u\,v^4\,w + 0.0162882\,v^5\,w + 2.22077\,u^4\,w^2 + 
   6.4639\,u^3\,v\,w^2 + 4.81333\,u^2\,v^2\,w^2 $\\&$+ 1.15478\,u\,v^3\,w^2 + 
   0.0555485\,v^4\,w^2 + 1.45911\,u^3\,w^3 + 3.06664\,u^2\,v\,w^3 + 
   1.35727\,u\,v^2\,w^3 + 0.0935349\,v^3\,w^3 + 0.522432\,u^2\,w^4 $\\&$+ 
   0.661835\,u\,v\,w^4 + 0.0871471\,v^2\,w^4 + 0.0804704\,u\,w^5 + 
   0.0286439\,v\,w^5 + 2.43619\,u^5\,y + 6.81471\,u^4\,v\,y + 
   5.57615\,u^3\,v^2\,y $\\&$+ 2.06187\,u^2\,v^3\,y + 0.29933\,u\,v^4\,y + 
   5.92205\,u^4\,w\,y + 12.266\,u^3\,v\,w\,y + 7.27261\,u^2\,v^2\,w\,y + 
   1.54143\,u\,v^3\,w\,y + 0.037268\,v^4\,w\,y $\\&$+ 5.83645\,u^3\,w^2\,y + 
   8.62368\,u^2\,v\,w^2\,y + 2.93576\,u\,v^2\,w^2\,y + 
   0.111143\,v^3\,w^2\,y + 2.7863\,u^2\,w^3\,y + 2.37933\,u\,v\,w^3\,y $\\&$+ 
   0.148969\,v^2\,w^3\,y + 0.53647\,u\,w^4\,y + 0.0809503\,v\,w^4\,y + 
   3.14287\,u^4\,y^2 + 5.2433\,u^3\,v\,y^2 + 2.89085\,u^2\,v^2\,y^2 + 
   0.557283\,u\,v^3\,y^2 $\\&$+ 6.20834\,u^3\,w\,y^2+7.3735\,u^2\,v\,w\,y^2 + 
   2.24663\,u\,v^2\,w\,y^2 + 0.034735\,v^3\,w\,y^2 + 4.45664\,u^2\,w^2\,y^2 + 
   2.95624\,u\,v\,w^2\,y^2 $\\&$+ 0.0805995\,v^2\,w^2\,y^2 + 
   1.14677\,u\,w^3\,y^2 + 0.0642921\,v\,w^3\,y^2 + 1.93839\,u^3\,y^3 + 
   2.12342\,u^2\,v\,y^3 + 0.61093\,u\,v^2\,y^3 $\\&$+ 2.79532\,u^2\,w\,y^3 + 
   1.66738\,u\,v\,w\,y^3 + 0.0148483\,v^2\,w\,y^3 + 1.08348\,u\,w^2\,y^3 + 
   0.020416\,v\,w^2\,y^3 + 0.630396\,u^2\,y^4 $\\&$+ 0.362335\,u\,v\,y^4 + 
   0.489119\,u\,w\,y^4 + 0.0022041\,v\,w\,y^4 + 0.0874933\,u\,y^5$
\end{tabular}
\end{table}

\begin{table}[t]
\squeezetable
\caption{
The five-loop series of
$\beta_v = m \partial v/\partial m$ for $M=3$.
Setting $\beta_v \equiv \sum_{l=0} \beta_{v,l}$, 
where $l$ is the number of loops, we report
$\beta_{v,l}$ up to $l=5$.
}
\label{tabMC3v}
\begin{tabular}{c|l}
\multicolumn{1}{c}{$l$}&
\multicolumn{1}{c}{$\beta_{v,l}$}\\
\tableline
0&$-v$\\
\hline 
1&$\case{3\,u\,v}{2} + v^2 + \case{v\,w}{2} + \case{2\,v\,y}{3}$\\
\hline 
2&$- \case{185\,u^2\,v}{216} - \case{412\,u\,v^2}{297} - 
   \case{1252\,v^3}{3267} - \case{77\,u\,v\,w}{108} - \case{47\,v^2\,w}{99} - 
   \case{23\,v\,w^2}{216} - \case{77\,u\,v\,y}{81} - 
   \case{400\,v^2\,y}{891} - \case{23\,v\,w\,y}{81} - \case{92\,v\,y^2}{729}$\\
\hline 
3&$  0.916668\,u^3\,v + 2.35093\,u^2\,v^2 + 1.40888\,u\,v^3 + 0.28295\,v^4 + 
   1.2163\,u^2\,v\,w + 1.79079\,u\,v^2\,w + 0.542156\,v^3\,w + 
   0.49351\,u\,v\,w^2 $\\&$+ 0.367154\,v^2\,w^2 + 0.0730773\,v\,w^3 + 
   1.62177\,u^2\,v\,y + 1.7731\,u\,v^2\,y + 0.508415\,v^3\,y + 
   1.31602\,u\,v\,w\,y + 0.746208\,v^2\,w\,y $\\&$+ 0.292309\,v\,w^2\,y + 
   0.63527\,u\,v\,y^2 + 0.346201\,v^2\,y^2 + 0.299507\,v\,w\,y^2 + 
   0.090449\,v\,y^3$\\
\hline 
4&$  -1.22868\,u^4\,v - 4.3522\,u^3\,v^2 - 4.33525\,u^2\,v^3-1.75357\,u\,v^4 - 
   0.270333\,v^5 - 2.26103\,u^3\,v\,w - 5.40209\,u^2\,v^2\,w - 
   3.5142\,u\,v^3\,w $\\&$- 0.712012\,v^4\,w - 1.53326\,u^2\,v\,w^2 - 
   2.40612\,u\,v^2\,w^2 - 0.798138\,v^3\,w^2 - 0.486101\,u\,v\,w^3 - 
   0.401462\,v^2\,w^3 - 0.06557\,v\,w^4 $\\&$- 3.0147\,u^3\,v\,y - 
   5.65057\,u^2\,v^2\,y - 3.27089\,u\,v^3\,y - 0.650693\,v^4\,y - 
   4.0887\,u^2\,v\,w\,y - 5.03042\,u\,v^2\,w\,y - 1.47787\,v^3\,w\,y $\\&$- 
   1.9444\,u\,v\,w^2\,y - 1.26296\,v^2\,w^2\,y - 0.349706\,v\,w^3\,y - 
   2.1004\,u^2\,v\,y^2 - 2.31802\,u\,v^2\,y^2 - 0.671001\,v^3\,y^2 - 
   2.02679\,u\,v\,w\,y^2 $\\&$- 1.16492\,v^2\,w\,y^2 - 0.550107\,v\,w^2\,y^2 - 
   0.616942\,u\,v\,y^3 - 0.350699\,v^2\,y^3 - 0.335525\,v\,w\,y^3 - 
   0.0754467\,v\,y^4$\\
\hline 
5&$ 1.97599\,u^5\,v + 9.00629\,u^4\,v^2 + 12.7239\,u^3\,v^3 + 
   7.96583\,u^2\,v^4 + 2.46634\,u\,v^5 + 0.312556\,v^6 + 4.69417\,u^4\,v\,w + 
   15.6644\,u^3\,v^2\,w $\\&$+ 16.0384\,u^2\,v^3\,w + 6.69086\,u\,v^4\,w + 
   1.04399\,v^5\,w + 4.50574\,u^3\,v\,w^2 + 11.0238\,u^2\,v^2\,w^2 + 
   7.53313\,u\,v^3\,w^2 + 1.60193\,v^4\,w^2 $\\&$+ 2.28761\,u^2\,v\,w^3 + 
   3.82594\,u\,v^2\,w^3 + 1.35045\,v^3\,w^3 + 0.642513\,u\,v\,w^4 + 
   0.562428\,v^2\,w^4 + 0.0804704\,v\,w^5 + 6.25889\,u^4\,v\,y $\\&$+ 
   16.8745\,u^3\,v^2\,y + 15.2053\,u^2\,v^3\,y + 6.1052\,u\,v^4\,y + 
   0.947895\,v^5\,y + 12.0153\,u^3\,v\,w\,y + 23.5313\,u^2\,v^2\,w\,y + 
   14.2714\,u\,v^3\,w\,y $\\&$+ 2.9145\,v^4\,w\,y + 9.15043\,u^2\,v\,w^2\,y + 
   12.2315\,u\,v^2\,w^2\,y + 3.8419\,v^3\,w^2\,y + 3.42673\,u\,v\,w^3\,y + 
   2.40324\,v^2\,w^3\,y + 0.53647\,v\,w^4\,y $\\&$+ 6.36312\,u^3\,v\,y^2 + 
   11.045\,u^2\,v^2\,y^2 + 6.4832\,u\,v^3\,y^2 + 1.31809\,v^4\,y^2 + 
   9.71176\,u^2\,v\,w\,y^2+11.5283\,u\,v^2\,w\,y^2+ 3.48488\,v^3\,w\,y^2$\\&$+ 
   5.47297\,u\,v\,w^2\,y^2 + 3.4119\,v^2\,w^2\,y^2 + 1.14677\,v\,w^3\,y^2 + 
   3.01986\,u^2\,v\,y^3 + 3.47764\,u\,v^2\,y^3 + 1.04671\,v^3\,y^3 + 
   3.42367\,u\,v\,w\,y^3 $\\&$+ 2.05758\,v^2\,w\,y^3 + 1.08348\,v\,w^2\,y^3 + 
   0.771673\,u\,v\,y^4 + 0.461776\,v^2\,y^4 + 0.489119\,v\,w\,y^4 + 
   0.0874933\,v\,y^5$
\end{tabular}
\end{table}

\begin{table}[t]
\squeezetable
\caption{
The five-loop series of
$\beta_w = m \partial w/\partial m$ for $M=3$.
Setting $\beta_w \equiv \sum_{l=0} \beta_{w,l}$, 
where $l$ is the number of loops, we report
$\beta_{w,l}$ up to $l=5$.
}
\label{tabMC3w}
\begin{tabular}{c|l}
\multicolumn{1}{c}{$l$}&
\multicolumn{1}{c}{$\beta_{w,l}$}\\
\tableline
0&$-w$\\
\hline 
1&$ \case{3\,u\,w}{2} + \case{4\,v\,w}{11} + w^2 + \case{2\,w\,y}{3}$\\
\hline 
2&$- \case{185\,u^2\,w}{216} - \case{223\,u\,v\,w}{297} - 
   \case{244\,v^2\,w}{3267} - \case{131\,u\,w^2}{108} - 
   \case{47\,v\,w^2}{99} - \case{95\,w^3}{216} - \case{77\,u\,w\,y}{81} - 
   \case{184\,v\,w\,y}{891} - \case{50\,w^2\,y}{81} - \case{92\,w\,y^2}{729}$\\
\hline 
3&$  0.916668\,u^3\,w + 1.37496\,u^2\,v\,w + 0.362132\,u\,v^2\,w + 
   0.042103\,v^3\,w + 1.98317\,u^2\,w^2 + 1.74277\,u\,v\,w^2 + 
   0.271146\,v^2\,w^2$\\&$+1.48661\,u\,w^3 + 0.676067\,v\,w^3+ 0.389923\,w^4 + 
   1.62177\,u^2\,w\,y + 0.939981\,u\,v\,w\,y + 0.164277\,v^2\,w\,y + 
   2.27978\,u\,w^2\,y $\\&$+ 0.732074\,v\,w^2\,y + 0.857364\,w^3\,y + 
   0.63527\,u\,w\,y^2 + 0.217823\,v\,w\,y^2 + 0.467389\,w^2\,y^2 + 
   0.090449\,w\,y^3$\\
\hline 
4&$ -1.22868\,u^4\,w - 2.66348\,u^3\,v\,w - 1.4268\,u^2\,v^2\,w - 
   0.347251\,u\,v^3\,w - 0.0326696\,v^4\,w - 3.58788\,u^3\,w^2 - 
   5.25808\,u^2\,v\,w^2 $\\&$- 1.93572\,u\,v^2\,w^2 - 0.244281\,v^3\,w^2 - 
   4.09897\,u^2\,w^3 - 3.9245\,u\,v\,w^3 - 0.759482\,v^2\,w^3 - 
   2.17101\,u\,w^4 - 1.07631\,v\,w^4 $\\&$-0.447316\,w^5 - 3.0147\,u^3\,w\,y - 
   3.33336\,u^2\,v\,w\,y - 1.18975\,u\,v^2\,w\,y - 0.154387\,v^3\,w\,y - 
   6.5664\,u^2\,w^2\,y - 4.83085\,u\,v\,w^2\,y $\\&$- 0.899584\,v^2\,w^2\,y - 
   5.02574\,u\,w^3\,y - 1.95255\,v\,w^3\,y - 1.34386\,w^4\,y $\\&$- 
   2.1004\,u^2\,w\,y^2 - 1.46007\,u\,v\,w\,y^2 - 0.286402\,v^2\,w\,y^2 - 
   3.1138\,u\,w^2\,y^2 - 1.15025\,v\,w^2\,y^2 - 1.2213\,w^3\,y^2 - 
   0.616942\,u\,w\,y^3 $\\&$- 0.244019\,v\,w\,y^3 - 0.476234\,w^2\,y^3 - 
   0.0754467\,w\,y^4$\\
\hline 
5&$  1.97599\,u^5\,w + 5.70624\,u^4\,v\,w + 4.72927\,u^3\,v^2\,w + 
   1.77266\,u^2\,v^3\,w + 0.357776\,u\,v^4\,w + 0.0327712\,v^5\,w + 
   7.28706\,u^4\,w^2 $\\&$+ 15.3337\,u^3\,v\,w^2 + 
   9.42002\,u^2\,v^2\,w^2 + 2.47241\,u\,v^3\,w^2 + 0.26882\,v^4\,w^2 + 
   11.2179\,u^3\,w^3 + 17.1513\,u^2\,v\,w^3 + 
   7.14394\,u\,v^2\,w^3 $\\&$+ 0.985494\,v^3\,w^3 + 8.985399\,u^2\,w^4 + 
   9.21597\,u\,v\,w^4 + 1.99704\,v^2\,w^4 + 3.72266\,u\,w^5 + 
   1.9646\,v\,w^5 + 0.633855\,w^6 $\\&$+ 6.25889\,u^4\,w\,y + 
   10.5195\,u^3\,v\,w\,y + 5.91865\,u^2\,v^2\,w\,y + 1.6148\,u\,v^3\,w\,y + 
   0.18753\,v^4\,w\,y + 18.525\,u^3\,w^2\,y + 22.759\,u^2\,v\,w^2\,y $\\&$+ 
   8.91572\,u\,v^2\,w^2\,y + 1.29356\,v^3\,w^2\,y + 21.5756\,u^2\,w^3\,y + 
   18.0042\,u\,v\,w^3\,y + 3.7067\,v^2\,w^3\,y + 11.6445\,u\,w^4\,y + 
   5.06192\,v\,w^4\,y $\\&$+ 2.43619\,w^5\,y + 6.36312\,u^3\,w\,y^2 + 
   7.15909\,u^2\,v\,w\,y^2 + 2.9287\,u\,v^2\,w\,y^2 + 0.453078\,v^3\,w\,y^2 + 
   14.4006\,u^2\,w^2\,y^2 $\\&$+ 11.226\,u\,v\,w^2\,y^2 + 
   2.42873\,v^2\,w^2\,y^2 + 11.4247\,u\,w^3\,y^2 + 4.69358\,v\,w^3\,y^2 + 
   3.14287\,w^4\,y^2 + 3.01986\,u^2\,w\,y^3  $\\&$+ 
   0.566387\,v^2\,w\,y^3 + 4.73169\,u\,w^2\,y^3 + 2.0427\,v\,w^2\,y^3 + 
   1.93839\,w^3\,y^3 + 0.771673\,u\,w\,y^4 + 0.355723\,v\,w\,y^4 + 
   0.630396\,w^2\,y^4 $\\&$+ 2.45617\,u\,v\,w\,y^3+ 0.0874933\,w\,y^5$
\end{tabular}
\end{table}

\begin{table}[t]
\squeezetable
\squeezetable
\caption{ 
The five-loop series of
$\beta_y = m \partial y/\partial m$ for $M=3$.
Setting $\beta_y \equiv \sum_{l=0} \beta_{y,l}$, 
where $l$ is the number of loops, we report
$\beta_{y,l}$ up to $l=5$.
}
\label{tabMC3y}
\begin{tabular}{c|l}
\multicolumn{1}{c}{$l$}&
\multicolumn{1}{c}{$\beta_{y,l}$}\\
\tableline
0&$-y$\\
\hline 
1&$  \case{9\,v\,w}{11} + \case{3\,u\,y}{2} + \case{12\,v\,y}{11} + 
   \case{3\,w\,y}{2} + y^2$\\
\hline 
2&$-\case{9\,u\,v\,w}{11} - \case{63\,v^2\,w}{121} - \case{27\,v\,w^2}{44} - 
   \case{185\,u^2\,y}{216} - \case{439\,u\,v\,y}{297} - 
   \case{1756\,v^2\,y}{3267} - \case{185\,u\,w\,y}{108} - 
   \case{185\,v\,w\,y}{99} $\\&$- \case{185\,w^2\,y}{216} - 
   \case{104\,u\,y^2}{81} - \case{832\,v\,y^2}{891} - 
   \case{104\,w\,y^2}{81} - \case{308\,y^3}{729}$\\
\hline 
3&$  1.25483\,u^2\,v\,w + 1.62749\,u\,v^2\,w + 0.44594\,v^3\,w + 
   1.84622\,u\,v\,w^2 + 1.1317\,v^2\,w^2 + 0.690211\,v\,w^3 + 
   0.916668\,u^3\,y $\\&$+ 2.49036\,u^2\,v\,y + 1.80879\,u\,v^2\,y + 
   0.438494\,v^3\,y + 2.75\,u^2\,w\,y + 5.96358\,u\,v\,w\,y + 
   2.36751\,v^2\,w\,y + 2.75\,u\,w^2\,y + 3.2512\,v\,w^2\,y $\\&$+ 
   0.91667\,w^3\,y + 2.133\,u^2\,y^2 + 3.1195\,u\,v\,y^2 + 
   1.13436\,v^2\,y^2 + 4.26599\,u\,w\,y^2 + 3.38291\,v\,w\,y^2 + 
   2.133\,w^2\,y^2 + 1.47806\,u\,y^3 $\\&$+1.07495\,v\,y^3 + 1.47806\,w\,y^3 + 
   0.35107\,y^4$\\
\hline 
4&$  -2.17122\,u^3\,v\,w - 4.35781\,u^2\,v^2\,w - 2.48256\,u\,v^3\,w - 
   0.48565\,v^4\,w - 4.77724\,u^2\,v\,w^2 - 6.00211\,u\,v^2\,w^2 - 
   1.72925\,v^3\,w^2 $\\&$- 3.62343\,u\,v\,w^3 - 2.2762\,v^2\,w^3 - 
   0.951639\,v\,w^4 - 1.22868\,u^4\,y - 4.59345\,u^3\,v\,y - 
   5.3004\,u^2\,v^2\,y - 2.55397\,u\,v^3\,y $\\&$- 0.464358\,v^4\,y - 
   4.91474\,u^3\,w\,y - 16.2997\,u^2\,v\,w\,y - 13.3566\,u\,v^2\,w\,y - 
   3.38345\,v^3\,w\,y - 7.37211\,u^2\,w^2\,y - 17.4996\,u\,v\,w^2\,y $\\&$- 
   7.45225\,v^2\,w^2\,y - 4.91474\,u\,w^3\,y - 6.09451\,v\,w^3\,y - 
   1.22868\,w^4\,y - 3.89927\,u^3\,y^2 - 9.02889\,u^2\,v\,y^2 - 
   6.52141\,u\,v^2\,y^2 $\\&$- 1.58095\,v^3\,y^2 - 11.6978\,u^2\,w\,y^2 - 
   19.0578\,u\,v\,w\,y^2 - 7.19018\,v^2\,w\,y^2 - 11.6978\,u\,w^2\,y^2 - 
   9.94972\,v\,w^2\,y^2 - 3.89927\,w^3\,y^2 $\\&$- 4.23696\,u^2\,y^3 - 
   6.1212\,u\,v\,y^3 - 2.22589\,v^2\,y^3 - 8.47392\,u\,w\,y^3 - 
   6.37921\,v\,w\,y^3 - 4.23696\,w^2\,y^3 $\\&$- 2.03309\,u\,y^4 - 
   1.47861\,v\,y^4 - 2.03309\,w\,y^4 - 0.376527\,y^5$\\
\hline 
5&$  4.2429179\,u^4\,v\,w + 11.7357\,u^3\,v^2\,w + 
   10.5256\,u^2\,v^3\,w + 4.10472\,u\,v^4\,w + 0.618728\,v^5\,w + 
   12.4807\,u^3\,v\,w^2 + 24.3706\,u^2\,v^2\,w^2 $\\&$+ 14.3671\,u\,v^3\,w^2 + 
   2.82095\,v^4\,w^2 + 14.2955\,u^2\,v\,w^3 + 18.2581\,u\,v^2\,w^3 + 
   5.42606\,v^3\,w^3 + 7.5579084\,u\,v\,w^4 + 4.86798\,v^2\,w^4 $\\&$+ 
   1.55088\,v\,w^5 + 1.97599\,u^5\,y + 9.47772\,u^4\,v\,y + 
   15.161\,u^3\,v^2\,y + 11.1288\,u^2\,v^3\,y + 4.00642\,u\,v^4\,y + 
   0.582751\,v^5\,y $\\&$+ 9.87996\,u^4\,w\,y + 44.5209\,u^3\,v\,w\,y + 
   56.6807\,u^2\,v^2\,w\,y + 28.7278\,u\,v^3\,w\,y + 5.3291\,v^4\,w\,y + 
   19.760\,u^3\,w^2\,y+71.398\,u^2\,v\,w^2\,y$\\&$+ 61.8264\,u\,v^2\,w^2\,y + 
   16.0162\,v^3\,w^2\,y + 19.760\,u^2\,w^3\,y + 49.2653\,u\,v\,w^3\,y + 
   21.8335\,v^2\,w^3\,y + 9.87996\,u\,w^4\,y + 12.6315\,v\,w^4\,y $\\&$+ 
   1.97599\,w^5\,y + 7.98749\,u^4\,y^2 + 25.6071\,u^3\,v\,y^2 + 
   28.2134\,u^2\,v^2\,y^2 + 13.5424\,u\,v^3\,y^2 + 2.46225\,v^4\,y^2 + 
   31.9499\,u^3\,w\,y^2$\\&$+80.4247\,u^2\,v\,w\,y^2+60.7793\,u\,v^2\,w\,y^2 + 
   14.9778\,v^3\,w\,y^2 + 47.9249\,u^2\,w^2\,y^2 + 82.7197\,u\,v\,w^2\,y^2 + 
   32.0629\,v^2\,w^2\,y^2 $\\&$+ 31.9499\,u\,w^3\,y^2 + 28.2945\,v\,w^3\,y^2 + 
   7.98749\,w^4\,y^2 + 11.9097\,u^3\,y^3 + 26.2915\,u^2\,v\,y^3 + 
   18.9292\,u\,v^2\,y^3 + 4.58889\,v^3\,y^3 $\\&$+ 35.729\,u^2\,w\,y^3 + 
   53.9545\,u\,v\,w\,y^3 + 19.9234\,v^2\,w\,y^3 + 35.729\,u\,w^2\,y^3 + 
   27.6713\,v\,w^2\,y^3 + 11.9097\,w^3\,y^3 + 8.72598\,u^2\,y^4 $\\&$+ 
   12.5648\,u\,v\,y^4 + 4.56903\,v^2\,y^4 + 17.452\,u\,w\,y^4 + 
   12.881\,v\,w\,y^4 + 8.72598\,w^2\,y^4 + 3.24652\,u\,y^5 + 2.3611\,v\,y^5 + 
   3.24652\,w\,y^5 $\\&$+ 0.495548\,y^6$
\end{tabular}
\end{table}

\begin{table}[t]
\squeezetable
\caption{ 
The five-loop series of
$\beta_u = m \partial u/\partial m$ for $M=2$.
Setting $\beta_u \equiv \sum_{l=0} \beta_{u,l}$, 
where $l$ is the number of loops, we report
$\beta_{u,l}$ up to $l=5$.
}
\label{tabMC2u}
\begin{tabular}{c|l}
\multicolumn{1}{c}{$l$}&
\multicolumn{1}{c}{$\beta_{u,l}$}\\
\tableline
0&$-u$\\
\hline
1&$u^2+\case{4}{5}u v+\case{1}{2} u w+ \case{1}{5} v w+ \case{2}{3} u y$\\
\hline
2&$-\case{95\,u^3}{216} - \case{20\,u^2\,v}{27} - \case{92\,u\,v^2}{675} - 
   \case{25\,u^2\,w}{54} - \case{41\,u\,v\,w}{90} - \case{2\,v^2\,w}{75} - 
   \case{23\,u\,w^2}{216} - \case{v\,w^2}{30} - \case{50\,u^2\,y}{81} - 
   \case{92\,u\,v\,y}{405} - \case{23\,u\,w\,y}{81} - \case{92\,u\,y^2}{729}$\\
\hline
3& $0.389923\,u^4 + 1.02884\,u^3\,v + 0.553406\,u^2\,v^2 + 0.101002\,u\,v^3 +  
   0.643023\,u^3\,w + 1.22281\,u^2\,v\,w + 0.444702\,u\,v^2\,w +  $\\&$
   0.025201\,v^3\,w + 0.35637\,u^2\,w^2 + 0.429803\,u\,v\,w^2 +  
   0.0577681\,v^2\,w^2 + 0.0730773\,u\,w^3 + 0.0279534\,v\,w^3 +  
   0.857364\,u^3\,y $\\&$+ 0.922343\,u^2\,v\,y + 0.252504\,u\,v^2\,y +  
   0.95032\,u^2\,w\,y + 0.646426\,u\,v\,w\,y + 0.028814\,v^2\,w\,y +  
   0.292309\,u\,w^2\,y + 0.03602\,v\,w^2\,y $\\&$+  0.467389\,u^2\,y^2 + 
   0.248819\,u\,v\,y^2 + 0.299507\,u\,w\,y^2 + 0.00460659\,v\,w\,y^2 +  
   0.090449\,u\,y^3$ \\
\hline
4& $-0.447316\,u^5 - 1.61263\,u^4\,v - 1.50485\,u^3\,v^2 -0.539552\,u^2\,v^3 - 
   0.0779026\,u\,v^4 - 1.0079\,u^4\,w - 2.88772\,u^3\,v\,w $\\&$ - 
   2.0261\,u^2\,v^2\,w - 0.444498\,u\,v^3\,w - 0.0187035\,v^4\,w - 
   0.885278\,u^3\,w^2 - 1.88177\,u^2\,v\,w^2 - 0.82283\,u\,v^2\,w^2 - 
   0.051806\,v^3\,w^2 $\\&$ - 0.374191\,u^2\,w^3 - 0.493586\,u\,v\,w^3 - 
   0.0658781\,v^2\,w^3 - 0.06557\,u\,w^4 - 0.0255368\,v\,w^4 - 
   1.34386\,u^4\,y - 2.50809\,u^3\,v\,y $\\&$ - 1.34888\,u^2\,v^2\,y - 
   0.259675\,u\,v^3\,y - 2.36074\,u^3\,w\,y - 3.28057\,u^2\,v\,w\,y - 
   1.0336\,u\,v^2\,w\,y - 0.031355\,v^3\,w\,y - 1.49676\,u^2\,w^2\,y $\\&$ - 
   1.2471\,u\,v\,w^2\,y - 0.06253\,v^2\,w^2\,y - 0.349706\,u\,w^3\,y - 
   0.0505394\,v\,w^3\,y - 1.2213\,u^3\,y^2 - 1.32086\,u^2\,v\,y^2 - 
   0.381274\,u\,v^2\,y^2 $\\&$ -1.5635\,u^2\,w\,y^2 - 0.966102\,u\,v\,w\,y^2 - 
   0.0164433\,v^2\,w\,y^2 - 0.550107\,u\,w^2\,y^2 - 0.0205542\,v\,w^2\,y^2 - 
   0.476234\,u^2\,y^3 $\\&$ - 0.274796\,u\,v\,y^3 - 0.335525\,u\,w\,y^3 - 
   0.0031877\,v\,w\,y^3 - 0.0754467\,u\,y^4$\\
\hline
5&$0.633855\,u^6 + 2.92343\,u^5\,v + 3.946\,u^4\,v^2 + 2.28569\,u^3\,v^3 + 
   0.683062\,u^2\,v^4 + 0.0863513\,u\,v^5 + 1.82714\,u^5\,w +  
   7.05664\,u^4\,v\,w $\\&$ + 7.67985\,u^3\,v^2\,w + 3.31184\,u^2\,v^3\,w +  
   0.600884\,u\,v^4\,w + 0.021403\,v^5\,w + 2.22077\,u^4\,w^2 +  
   6.8986\,u^3\,v\,w^2 + 5.54476\,u^2\,v^2\,w^2 $\\&$+ 1.44276\,u\,v^3\, w^2 + 
   0.0762973\,v^4\,w^2 + 1.45911\,u^3\,w^3 + 3.34723\,u^2\,v\,w^3 +  
   1.6199\,u\,v^2\,w^3 + 0.122458\,v^3\,w^3 + 0.522432\,u^2\,w^4 $\\&$ +  
   0.725783\,u\,v\,w^4 + 0.104985\,v^2\,w^4 + 0.0804704\,u\,w^5 +   
   0.0315083\,v\,w^5 + 2.43619\,u^5\,y + 6.57667\,u^4\,v\,y +  
   5.71423\,u^3\,v^2\,y $\\&$+ 2.27687\,u^2\,v^3\,y + 0.359797\,u\,v^4\,y +  
   5.92205\,u^4\,w\,y + 12.9278\,u^3\,v\,w\,y + 8.30663\,u^2\,v^2\,w\,y +  
   1.91539\,u\,v^3\,w\,y $\\&$+ 0.0515184\,v^4\,w\,y + 5.83645\,u^3\,w^2\,y +  
   9.38178\,u^2\,v\,w^2\,y + 3.48908\,u\,v^2\,w^2\,y +  
   0.146069\,v^3\,w^2\,y + 2.78631\,u^2\,w^3\,y $\\&$+ 2.60534\,u\,v\,w^3\,y + 
   0.17948\,v^2\,w^3\,y + 0.53647\,u\,w^4\,y + 0.089045\,v\,w^4\,y +  
   3.14287\,u^4\,y^2 + 5.50064\,u^3\,v\,y^2 + 3.27895\,u^2\,v^2\,y^2$\\&$ +  
   0.689359\,u\,v^3\,y^2 + 6.20834\,u^3\,w\,y^2 + 8.00363\,u^2\,v\,w\,y^2+
   2.6647\,u\,v^2\,w\,y^2 + 0.0459087\,v^3\,w\,y^2 + 4.45664\,u^2\,w^2\,y^2 $\\&$ + 
   3.23343\,u\,v\,w^2\,y^2 + 0.0973883\,v^2\,w^2\,y^2 +  
   1.14677\,u\,w^3\,y^2 + 0.0707213\,v\,w^3\,y^2 + 1.93839\,u^3\,y^3  + 
   2.30259\,u^2\,v\,y^3 $\\&$+ 0.723961\,u\,v^2\,y^3 + 2.79532\,u^2\,w\,y^3  + 
   1.82321\,u\,v\,w\,y^3 + 0.0179664\,v^2\,w\,y^3 + 1.08348\,u\,w^2\,y^3  + 
   0.022458\,v\,w^2\,y^3 $\\&$+ 0.630396\,u^2\,y^4 + 0.396144\,u\,v\,y^4 +  
   0.489119\,u\,w\,y^4 + 0.00242451\,v\,w\,y^4 + 0.0874933\,u\,y^5$ 
\end{tabular}
\end{table}

\begin{table}[t]
\squeezetable
\caption{
The five-loop series of
$\beta_v = m \partial v/\partial m$ for $M=2$.
Setting $\beta_v \equiv \sum_{l=0} \beta_{v,l}$, 
where $l$ is the number of loops, we report
$\beta_{v,l}$ up to $l=5$.
}
\label{tabMC2v}
\begin{tabular}{c|l}
\multicolumn{1}{c}{$l$}&
\multicolumn{1}{c}{$\beta_{v,l}$}\\
\tableline
0&$-v$\\
\hline 
1&$ v^2 +\case{3}{2} u v+\case{1}{2} v w+\case{2}{3} v y$\\
\hline 
2&$-\case{185\,u^2\,v}{216} - \case{181\,u\,v^2}{135} - \case{272\,v^3}{675} - 
    \case{77\,u\,v\,w}{108} - \case{47\,v^2\,w}{90} - \case{23\,v\,w^2}{216} - 
    \case{77\,u\,v\,y}{81} - \case{40\,v^2\,y}{81} - \case{23\,v\,w\,y}{81} - 
    \case{92\,v\,y^2}{729}$\\
\hline 
3&$  0.916668\,u^3\,v + 2.25286\,u^2\,v^2 + 1.44833\,u\,v^3 + 0.314917\,v^4 + 
    1.21633\,u^2\,v\,w + 1.9382\,u\,v^2\,w + 0.635652\,v^3\,w + 
    0.49351\,u\,v\,w^2 $\\&$+ 0.401926\,v^2\,w^2 + 0.0730773\,v\,w^3 + 
    1.62177\,u^2\,v\,y + 1.90822\,u\,v^2\,y + 0.593025\,v^3\,y + 
    1.31602\,u\,v\,w\,y + 0.815647\,v^2\,w\,y $\\&$+ 0.292309\,v\,w^2\,y + 
    0.63527\,u\,v\,y^2 + 0.378517\,v^2\,y^2 + 0.299507\,v\,w\,y^2 + 
    0.090449\,v\,y^3$\\
\hline 
4&$  -1.22868\,u^4\,v - 4.14839\,u^3\,v^2 - 4.30604\,u^2\,v^3 - 
    1.88967\,u\,v^4 - 0.317928\,v^5 - 2.26103\,u^3\,v\,w - 
    5.70167\,u^2\,v^2\,w - 4.000\,u\,v^3\,w $\\&$- 0.885211\,v^4\,w - 
    1.53326\,u^2\,v\,w^2 - 2.62769\,u\,v^2\,w^2 - 0.952554\,v^3\,w^2 - 
    0.486101\,u\,v\,w^3 - 0.440264\,v^2\,w^3 - 0.06557\,v\,w^4 $\\&$- 
    3.0147\,u^3\,v\,y - 5.8948\,u^2\,v^2\,y - 3.71358\,u\,v^3\,y - 
    0.807636\,v^4\,y - 4.0887\,u^2\,v\,w\,y - 5.48266\,u\,v^2\,w\,y - 
    1.7649\,v^3\,w\,y $\\&$- 1.9444\,u\,v\,w^2\,y - 1.38387\,v^2\,w^2\,y - 
    0.349706\,v\,w^3\,y - 2.1004\,u^2\,v\,y^2 - 2.53023\,u\,v^2\,y^2 - 
    0.801818\,v^3\,y^2 - 2.02679\,u\,v\,w\,y^2 $\\&$- 1.27603\,v^2\,w\,y^2 - 
    0.550107\,v\,w^2\,y^2 - 0.616942\,u\,v\,y^3 - 0.384175\,v^2\,y^3 - 
    0.335525\,v\,w\,y^3 - 0.0754467\,v\,y^4$\\
\hline 
5&$  1.97599\,u^5\,v + 8.54783\,u^4\,v^2 + 12.3856\,u^3\,v^3 + 
    8.36256\,u^2\,v^4 + 2.82533\,u\,v^5 + 0.391102\,v^6 + 4.69417\,u^4\,v\,w + 
    16.2955\,u^3\,v^2\,w $\\&$+ 17.8673\,u^2\,v^3\,w + 8.13124\,u\,v^4\,w + 
    1.38643\,v^5\,w + 4.50574\,u^3\,v\,w^2 + 11.9507\,u^2\,v^2\,w^2 + 
    8.92856\,u\,v^3\,w^2 + 2.07519\,v^4\,w^2 $\\&$+ 2.28761\,u^2\,v\,w^3 + 
    4.19835\,u\,v^2\,w^3 + 1.62409\,v^3\,w^3 + 0.642513\,u\,v\,w^4 + 
    0.617553\,v^2\,w^4 + 0.0804704\,v\,w^5 + 6.25889\,u^4\,v\,y $\\&$+ 
    17.3148\,u^3\,v^2\,y + 16.873\,u^2\,v^3\,y + 7.4013\,u\,v^4\,y + 
    1.25617\,v^5\,y + 12.0153\,u^3\,v\,w\,y + 25.4165\,u^2\,v^2\,w\,y + 
    16.8835\,u\,v^3\,w\,y $\\&$+3.77286\,v^4\,w\,y + 9.15043\,u^2\,v\,w^2\,y + 
    13.414\,u\,v^2\,w^2\,y + 4.61743\,v^3\,w^2\,y + 3.42673\,u\,v\,w^3\,y + 
    2.6376\,v^2\,w^3\,y + 0.53647\,v\,w^4\,y $\\&$+ 6.36312\,u^3\,v\,y^2 + 
    11.940\,u^2\,v^2\,y^2 + 7.66554\,u\,v^3\,y^2 + 1.70545\,v^4\,y^2 + 
    9.71176\,u^2\,v\,w\,y^2 + 12.639\,u\,v^2\,w\,y^2 + 4.18792\,v^3\,w\,y^2 $\\&$+ 
    5.47297\,u\,v\,w^2\,y^2 + 3.74384\,v^2\,w^2\,y^2 + 1.14677\,v\,w^3\,y^2 + 
    3.01986\,u^2\,v\,y^3 + 3.81205\,u\,v^2\,y^3 + 1.2579\,v^3\,y^3 + 
    3.42367\,u\,v\,w\,y^3 $\\&$+ 2.25788\,v^2\,w\,y^3 + 1.08348\,v\,w^2\,y^3 + 
    0.771673\,u\,v\,y^4 + 0.506741\,v^2\,y^4 + 0.489119\,v\,w\,y^4 + 
    0.0874933\,v\,y^5$
\end{tabular}
\end{table}

\begin{table}[t]
\squeezetable
\caption{
The five-loop series of
$\beta_w = m \partial w/\partial m$ for $M=2$.
Setting $\beta_w \equiv \sum_{l=0} \beta_{w,l}$, 
where $l$ is the number of loops, we report
$\beta_{w,l}$ up to $l=5$.
}
\label{tabMC2w}
\begin{tabular}{c|l}
\multicolumn{1}{c}{$l$}&
\multicolumn{1}{c}{$\beta_{w,l}$}\\
\tableline
0&$-w$\\
\hline 
1&$\case{3\,u\,w}{2} + \case{2\,v\,w}{5} + w^2 + \case{2\,w\,y}{3}$\\
\hline 
2&$-\case{185\,u^2\,w}{216} - \case{20\,u\,v\,w}{27} - 
    \case{56\,v^2\,w}{675} - \case{131\,u\,w^2}{108} - \case{47\,v\,w^2}{90} - 
    \case{95\,w^3}{216} - \case{77\,u\,w\,y}{81} - \case{92\,v\,w\,y}{405} - 
    \case{50\,w^2\,y}{81} - \case{92\,w\,y^2}{729}$\\
\hline 
3&$0.916668\,u^3\,w + 1.33266\,u^2\,v\,w + 0.406785\,u\,v^2\,w + 
   0.0505995\,v^3\,w + 1.98317\,u^2\,w^2 + 1.90301\,u\,v\,w^2 + 
   0.324868\,v^2\,w^2 $\\&$+1.48661\,u\,w^3+0.743674\,v\,w^3 + 0.389923\,w^4 + 
   1.62177\,u^2\,w\,y  + 1.04026\,u\,v\,w\,y + 0.194876\,v^2\,w\,y + 
   2.27978\,u\,w^2\,y $\\&$+ 0.805282\,v\,w^2\,y + 0.857364\,w^3\,y + 
    0.63527\,u\,w\,y^2 + 0.239606\,v\,w\,y^2 + 0.467389\,w^2\,y^2 + 
    0.090449\,w\,y^3$\\
\hline 
4&$  -1.22868\,u^4\,w - 2.55616\,u^3\,v\,w - 1.47403\,u^2\,v^2\,w - 
   0.384865\,u\,v^3\,w  - 0.0404956\,v^4\,w - 3.58788\,u^3\,w^2 - 
   5.59606\,u^2\,v\,w^2 $\\&$- 2.25333\,u\,v^2\,w^2 - 0.310265\,v^3\,w^2 - 
   4.09897\,u^2\,w^3 - 4.33198\,u\,v\,w^3 - 0.921371\,v^2\,w^3 -  
   2.17101\,u\,w^4 - 1.18394\,v\,w^4 - $\\&$0.447316\,w^5 -3.0147\,u^3\,w\,y - 
   3.53479\,u^2\,v\,w\,y - 1.37253\,u\,v^2\,w\,y  - 0.196966\,v^3\,w\,y - 
   6.5664\,u^2\,w^2\,y - 5.33631\,u\,v\,w^2\,y $\\&$- 1.09183\,v^2\,w^2\,y - 
   5.02574\,u\,w^3\,y - 2.1478\,v\,w^3\,y - 1.34386\,w^4\,y -  
   2.1004\,u^2\,w\,y^2 - 1.6112\,u\,v\,w\,y^2 - 0.348387\,v^2\,w\,y^2 $\\&$- 
   3.1138\,u\,w^2\,y^2 - 1.26528\,v\,w^2\,y^2 - 1.2213\,w^3\,y^2 -  
   0.616942\,u\,w\,y^3 - 0.26842\,v\,w\,y^3 - 0.476234\,w^2\,y^3 -  
   0.0754467\,w\,y^4$\\ 
\hline 
5&$1.97599 \,u^5\,w + 5.43635\,u^4\,v\,w + 4.70998\,u^3\,v^2\,w + 
   1.94106 \,u^2\,v^3\,w + 0.434203\,u\,v^4\,w + 0.0435452\,v^5\,w + 
   7.28706\,u^4\,w^2 $\\&$+ 16.0529\,u^3\,v\,w^2 + 
   10.7128\,u^2\,v^2\,w^2 + 3.09443\,u\,v^3\,w^2 + 0.3684\,v^4\,w^2 + 
   11.2180\,u^3\,w^3 + 18.7355\,u^2\,v\,w^3 +
   8.61411\,u\,v^2\,w^3 $\\&$ + 1.30212\,v^3\,w^3 + 8.9854\,u^2\,w^4 + 
   10.1884\,u\,v\,w^4 + 2.4288\,v^2\,w^4 + 3.72266\,u\,w^5 +  
   2.16106\,v\,w^5 + 0.63385\,w^6 + 6.25889\,u^4\,w\,y $\\&$ + 
   10.9184\,u^3\,v\,w\,y + 6.74643\,u^2\,v^2\,w\,y + 2.0261\,u\,v^3\,w\,y  + 
   0.25676\,v^4\,w\,y + 18.5254\,u^3\,w^2\,y + 24.8502\,u^2\,v\,w^2\,y $\\&$+ 
   10.7457\,u\,v^2\,w^2\,y  + 1.7078\,v^3\,w^2\,y + 21.5756\,u^2\,w^3\,y + 
   19.9199\,u\,v\,w^3\,y +  4.50834\,v^2\,w^3\,y + 11.6445\,u\,w^4\,y + 
   5.56811\,v\,w^4\,y $\\&$+ 2.4362\,w^5\,y + 6.3631\,u^3\,w\,y^2 + 
   7.8398\,u^2\,v\,w\,y^2 + 3.53015\,u\,v^2\,w\,y^2 + 0.59754\,v^3\,w\ ,y^2 + 
   14.401\,u^2\,w^2\,y^2 + 12.417\,u\,v\,w^2\,y^2 $\\&$+ 
   2.95262\,v^2\,w^2\,y^2 + 11.4247\,u\,w^3\,y^2 + 5.16294\,v \,w^3\,y^2 + 
   3.14287\,w^4\,y^2 + 3.01986\,u^2\,w\,y^3 + 2.71417\,u\,v\, w\,y^3 + 
   0.68803\,v^2\,w\,y^3 $\\&$+ 4.73169\,u\,w^2\,y^3 + 2.24697\,v\,w^2\,y^3 + 
   1.93839\,w^3\,y^3 + 0.771673\,u\,w\,y^4 + 0.391 295\,v\,w\,y^4 + 
   0.630396\,w^2\,y^4 + 0.0874933\,w\,y^5$
\end{tabular}
\end{table}

\begin{table}[t]
\squeezetable
\squeezetable
\caption{ 
The five-loop series of
$\beta_y = m \partial y/\partial m$ for $M=2$.
Setting $\beta_u \equiv \sum_{l=0} \beta_{y,l}$, 
where $l$ is the number of loops, we report
$\beta_{y,l}$ up to $l=5$.
}
\label{tabMC2y}
\begin{tabular}{c|l}
\multicolumn{1}{c}{$l$}&
\multicolumn{1}{c}{$\beta_{y,l}$}\\
\tableline
0&$-y$\\ 
\hline 
1&$ \case{9\,v\,w}{10} + \case{3\,u\,y}{2} + \case{6\,v\,y}{5} + 
    \case{3\,w\,y}{2} + y^2$\\
\hline 
2&$-\case{9\,u\,v\,w}{10} - \case{3\,v^2\,w}{5} - \case{27\,v\,w^2}{40} - 
    \case{185\,u^2\,y}{216} - \case{208\,u\,v\,y}{135} - 
    \case{416\,v^2\,y}{675} - \case{185\,u\,w\,y}{108} - 
    \case{37\,v\,w\,y}{18} $\\&$- \case{185\,w^2\,y}{216} - 
    \case{104\,u\,y^2}{81} - \case{416\,v\,y^2}{405} - 
    \case{104\,w\,y^2}{81} - \case{308\,y^3}{729}$\\
\hline 
3&$   1.38031\,u^2\,v\,w + 1.8544\,u\,v^2\,w + 0.559618\,v^3\,w + 
    2.03085\,u\,v\,w^2 + 1.36855\,v^2\,w^2 + 0.759232\,v\,w^3 + 
   0.916668\,u^3\,y $\\&$+ 2.5596\,u^2\,v\,y + 2.05514\,u\,v^2\,y + 
    0.548037\,v^3\,y + 2.75\,u^2\,w\,y + 6.52715\,u\,v\,w\,y + 
   2.86347\,v^2\,w\,y + 2.75\,u\,w^2\,y + 3.57632\,v\,w^2\,y $\\&$+ 
   0.916668\,w^3\,y + 2.133\,u^2\,y^2 + 3.42523\,u\,v\,y^2 + 
   1.37009\,v^2\,y^2 + 4.26599\,u\,w\,y^2 + 3.7212\,v\,w\,y^2 $\\&$+ 
    2.133\,w^2\,y^2 + 1.47806\,u\,y^3 + 1.18245\,v\,y^3 + 1.47806\,w\,y^3 + 
    0.35107\,y^4$\\
\hline 
4&$  -2.38834\,u^3\,v\,w - 4.91662\,u^2\,v^2\,w - 3.05722\,u\,v^3\,w - 
   0.652482\,v^4\,w - 5.25496\,u^2\,v\,w^2 - 7.19501\,u\,v^2\,w^2 -  
   2.2724\,v^3\,w^2 $\\&$- 3.98577\,u\,v\,w^3 - 2.75879\,v^2\,w^3 - 
   1.0468\,v\,w^4  - 1.22868\,u^4\,y - 
     4.67913\,u^3\,v\,y - 5.84436\,u^2\,v^2\,y - 3.1024\,u\,v^3\,y 
     $\\&$- 4.91474\,u^3\,w\,y - 17.653\,u^2\,v\,w\,y - 
    15.9343\,u\,v^2\,w\,y - 4.43201\,v^3\,w\,y - 7.37211\,u^2\,w^2\,y - 
   19.2778\,u\,v\,w^2\,y - 9.03619\,v^2\,w^2\,y $\\&$- 4.91474\,u\,w^3\,y - 
   6.70396\,v\,w^3\,y - 1.22868\,w^4\,y - 3.89927\,u^3\,y^2 - 
   9.7406\,u^2\,v\,y^2 - 7.756\,u\,v^2\,y^2 - 2.06827\,v^3\,y^2 $\\&$- 
   11.6978\,u^2\,w\,y^2 - 21.0095\,u\,v\,w\,y^2 - 8.71866\,v^2\,w\,y^2 - 
   11.6978\,u\,w^2\,y^2 - 10.9447\,v\,w^2\,y^2 - 3.89927\,w^3\,y^2 - 
   4.23696\,u^2\,y^3 $\\&$- 6.74859\,u\,v\,y^3 - 2.69944\,v^2\,y^3 - 
   8.47392\,u\,w\,y^3 - 7.01714\,v\,w\,y^3 - 4.23696\,w^2\,y^3 - 
   2.03309\,u\,y^4 - 1.62647\,v\,y^4$\\&$ - 2.03309\,w\,y^4 - 0.376527\,y^5
  -0.62048\,v^4\,y$\\
\hline 
5&$ 4.6672\,u^4\,v\,w + 13.1278\,u^3\,v^2\,w + 
   12.6693\,u^2\,v^3\,w + 5.41225\,u\,v^4\,w + 0.895128\,v^5\,w + 
   13.7288\,u^3\,v\,w^2 + 28.8549\,u^2\,v^2\,w^2$\\&$ + 18.6536\,u\,v^3\,w^2 + 
   4.02044\,v^4\,w^2 + 15.7251\,u^2\,v\,w^3 + 22.1012\,u\,v^2\,w^3 + 
   7.22562\,v^3\,w^3 + 8.3137\,u\,v\,w^4 + 5.90266\,v^2\,w^4 $\\&$+ 
   1.70597\,v\,w^5 + 1.97599\,u^5\,y + 9.58498\,u^4\,v\,y + 
   16.3791\,u^3\,v^2\,y + 13.2026\,u^2\,v^3\,y + 5.24509\,u\,v^4\,y + 
   0.839214\,v^5\,y $\\&$+ 9.87996\,u^4\,w\,y + 47.8346\,u^3\,v\,w\,y + 
   66.4888\,u^2\,v^2\,w\,y + 37.149\,u\,v^3\,w\,y + 7.5788\,v^4\,w\,y + 
   19.7599\,u^3\,w^2\,y+78.284\,u^2\,v\,w^2\,y $\\&$+74.7963\,u\,v^2\,w^2\,y+ 
   21.317\,v^3\,w^2\,y + 19.7599\,u^2\,w^3\,y + 54.3088\,u\,v\,w^3\,y + 
   26.4953\,v^2\,w^3\,y + 9.87996\,u\,w^4\,y + 13.8947\,v\,w^4\,y $\\&$+ 
   1.97599\,w^5\,y + 7.98749\,u^4\,y^2 + 27.2985\,u^3\,v\,y^2 + 
   33.0066\,u^2\,v^2\,y^2 + 17.4836\,u\,v^3\,y^2 + 3.49673\,v^4\,y^2 + 
   31.9499\,u^3\,w\,y^2$\\&$+88.0974\,u^2\,v\,w\,y^2+73.5089\,u\,v^2\,w\,y^2 + 
   19.9292\,v^3\,w\,y^2 + 47.9249\,u^2\,w^2\,y^2 + 91.2671\,u\,v\,w^2\,y^2 + 
   38.9216\,v^2\,w^2\,y^2 $\\&$+ 31.9499\,u\,w^3\,y^2 + 31.124\,v\,w^3\,y^2 + 
   7.98749\,w^4\,y^2 + 11.9097\,u^3\,y^3 + 28.8082\,u^2\,v\,y^3 + 
   22.8905\,u\,v^2\,y^3 + 6.10414\,v^3\,y^3 $\\&$+ 35.729\,u^2\,w\,y^3 + 
   59.5478\,u\,v\,w\,y^3 + 24.1885\,v^2\,w\,y^3 + 35.729\,u\,w^2\,y^3 + 
   30.4384\,v\,w^2\,y^3 + 11.9097\,w^3\,y^3 + 8.72598\,u^2\,y^4 $\\&$+ 
   13.8681\,u\,v\,y^4 + 5.54723\,v^2\,y^4 + 17.452\,u\,w\,y^4 + 
   14.1691\,v\,w\,y^4 + 8.72598\,w^2\,y^4 + 3.24652\,u\,y^5 + 
   2.59721\,v\,y^5 + 3.24652\,w\,y^5 $\\&$+ 0.495548\,y^6$
\end{tabular}
\end{table}

\begin{table}[tbp]
\squeezetable
\caption{
The coefficients $e^{U}_{ij}$, cf. Eq.~(\ref{yus}).
}
\label{tu}
\begin{tabular}{cl}
\multicolumn{1}{c}{$i,j$}&
\multicolumn{1}{c}{$(N+8)^i e^{U}_{ij}$}\\
\tableline 
3,0&$-18.3128 - 3.43328\,N + 0.216746\,N^2$\\
2,1&$-9.15642 - 0.17027\,N$\\
1,2&$-1.17334$\\
0,3&$-0.0443103$\\
\hline
4,0&$140.799 + 37.5734\,N + 1.03627\,N^2 + 0.0943426\,N^3$\\
3,1&$93.8662 + 5.52597\,N - 0.0781363\,N^2$\\
2,2&$18.4511 - 0.0667897\,N$\\
1,3&$1.40677$\\
0,4&$0.0395196$\\
\hline
5,0&$-1340.07 - 416.717\,N - 17.6226\,N^2 + 0.911281\,N^3 + 0.0508337\,N^4$\\
4,1&$-1116.73 - 98.6844\,N + 1.52723\,N^2 - 0.0301952\,N^3$\\
3,2&$-298.289 - 2.54766\,N - 0.0492195\,N^2$\\
2,3&$-35.2065 + 0.140508\,N$\\
1,4&$-1.98589$\\
0,5&$-0.0444004$\\
\hline
6,0&$15651.3 + 5665.65\,N + 433.687\,N^2 + 1.06755\,N^3 + 0.679106\,N^4 +  0.031393\,N^5$\\
5,1&$15651.3 + 1935.63\,N + 8.74297\,N^2 + 0.581411\,N^3 - 0.00927903\,N^4$\\
4,2&$5294.38 + 134.776\,N - 1.13059\,N^2 - 0.038664\,N^3$\\
3,3&$848.418 - 2.51108\,N + 0.0323227\,N^2$\\
2,4&$72.4799 - 0.318348\,N$\\
1,5&$3.24291$\\
0,6&$0.0603632$
\end{tabular}
\end{table}

\begin{table}[tbp]
\squeezetable
\caption{
The coefficients $e^{T}_{ij}$, cf. Eq.~(\ref{yts}).
}
\label{tt}
\begin{tabular}{cl}
\multicolumn{1}{c}{$i,j$}&
\multicolumn{1}{c}{$(N+8)^i e^{T}_{ij}$}\\
\tableline 
2,0&$-18.3128 - 3.43328\,N + 0.216746\,N^2$\\
1,1&$-3.09273 + 0.216746\,N$\\
0,2&$-0.0337239$\\
\hline
3,0&$140.799 + 37.5734\,N + 1.03627\,N^2 + 0.0943426\,N^3$\\
2,1&$39.0459 + 1.53797\,N + 0.12579\,N^2$\\
1,2&$2.66843 + 0.0769829\,N$\\
0,3&$0.0716893$\\
\hline
4,0&$-1340.07 - 416.717\,N - 17.6226\,N^2 + 0.911281\,N^3 + 0.0508337\,N^4$\\
3,1&$-497.159 - 32.4255\,N + 1.57919\,N^2 + 0.0847229\,N^3$\\
2,2&$-53.6464 + 0.443225\,N + 0.062623\,N^2$\\
1,3&$-2.39176 + 0.0256721\,N$\\
0,4&$-0.0421612$\\
\hline
5,0&$15651.3 + 5665.65\,N + 433.687\,N^2 + 1.06755\,N^3 + 0.679106\,N^4 +  0.031393\,N^5$\\
4,1&$7460.04 + 849.888\,N + 0.972279\,N^2 + 1.37677\,N^3 + 0.062786\,N^4$\\
3,2&$1175.99 + 25.2714\,N + 1.12305\,N^2 + 0.0567755\,N^3$\\
2,3&$88.2226 + 0.60426\,N + 0.0297911\,N^2$\\
1,4&$3.53359 + 0.0136874\,N$\\
0,5&$0.0607723$
\end{tabular}
\end{table}


\begin{references}
\bibitem{CPV-unpub}
We have extended the perturbative series for the isotropic RAM to 
five loops in the fixed-dimension massive zero-momentum scheme. 
The analysis does not find evidence of stable fixed points.
\bibitem{foot-DyAl2}
DyAl$_2$ is a ferromagnet with strong crystalline anisotropy,
[100] being the easy axis for magnetization. Thus, we expect this system to
show a first-order transition, see Ref.~\cite{PV-r}. Experiments
\cite{GSMA-90}, however, apparently observe a Heisenberg second-order 
transition. This means that the first-order transition is very weak and that,
in the range of temperatures investigated experimentally, DyAl$_2$ 
can be effectively considered as a Heisenberg system. It makes therefore sense 
to compare the experimental results for 
(Dy$_x$Y$_{1-x}$)Al$_2$ with the theoretical predictions for the crossover 
close to a Heisenberg transition.
\bibitem{def-PB}
Given an expansion $\sum c_n g^n$, we define the 
Borel-Leroy transform $B(t) = c_n t^n/\Gamma(n+b+1)$, 
and a Pad\'e approximant $P(t)$ to $B(t)$. Then, a resummed expression 
is obtained by considering $\int_0^\infty dt\, t^b e^{-t} P(t)$. 
The parameter $b$ may be chosen at will.
\bibitem{foot-v0-inf} Indeed, for $D=0$, the potential appearing in the 
Hamiltonian (\ref{heffRCAM}) can be written as 
$V(\varphi) = {1\over 4!} v_0 (\vec{\varphi}^2 + 6 r/v_0)^2 + 
{1\over 4!} y_0 \sum_a \varphi_a^4$. Fixed-length spins are recovered
by taking the limit $v_0\to +\infty$, $r\to -\infty$ keeping 
$r/v_0 = - 1/6$.
\bibitem{foot-XY}
For $K\to-\infty$ the variables $\vec{s}$ are constrained to assume values 
$(\pm1,0)$ and $(0,\pm1)$. Thus, we can define $s_1 = {1\over2}(\sigma+\tau)$
and $s_2 = {1\over2}(\sigma-\tau)$, where $\sigma$ and $\tau$ are 
unconstrained Ising variables. The Hamiltonian becomes that of two 
uncoupled Ising models, since $\vec{s}_x\cdot\vec{s}_y = {1\over2} 
 (\sigma_x\sigma_y + \tau_x\tau_y)$.
\end{references}
\end{document}